\documentclass[aps,prb,twocolumn,preprintnumbers,amsmath,amssymb,superscriptaddress]{revtex4-1}
\pdfoutput=1

\usepackage{graphicx}
\usepackage{bm}
\usepackage[usenames,dvipsnames]{xcolor}
\usepackage{hyperref}
\usepackage{pstricks}
\usepackage[tight]{subfigure}
\usepackage{verbatim}
\usepackage{units}
\usepackage{multirow}
\usepackage{enumitem}
\usepackage[rightcaption]{sidecap}

\usepackage[english]{babel}
\renewcommand{\vec}[1]{\mathbf{#1}}

\renewcommand{\vec}[1]{\bm{#1}}
\newcommand{\vecG}[1]{\bm{#1}}

\newcommand{\bra}[1]{\langle #1|}
\newcommand{\ket}[1]{|#1 \rangle}

\newcommand{\Ham}{\mathcal{H}}
\newcommand{\prob}{\mathcal{P}}
\newcommand{\kcL}{\mathcal{L}}
\newcommand{\ZT}{\textrm{ZT}}

\newcommand{\aW}{\mathcal{W}}
\newcommand{\aV}{\mathcal{V}}
\newcommand{\aT}{\mathcal{T}}

\newcommand{\aWW}{\mathbb{W}}
\newcommand{\aVV}{\mathbb{V}}

\usepackage[normalem]{ulem}

\usepackage{hyperref}
\hypersetup{colorlinks=true,linktoc=all,linkcolor=blue,citecolor=blue}

\begin{document}




\title{The effect of magnetic anisotropy on spin-dependent thermoelectric effects in nanoscopic systems}

\author{Maciej Misiorny}
 \email{misiorny@amu.edu.pl}
\affiliation{Peter Gr{\"u}nberg Institut PGI-2, Forschungszentrum J{\"u}lich, 52425 J{\"u}lich,  Germany}
\affiliation{JARA\,--\,Fundamentals of Future Information Technology, 52425 J{\"u}lich,  Germany}
\affiliation{Faculty of Physics, Adam Mickiewicz University, 61-614 Pozna\'{n}, Poland}

\author{J\'{o}zef Barna\'{s}}
\affiliation{Faculty of Physics, Adam Mickiewicz University, 61-614 Pozna\'{n}, Poland}
\affiliation{Institute of Molecular Physics, Polish Academy of Sciences, 60-179 Pozna\'{n}, Poland}

\date{\today}

\begin{abstract}
Conventional and spin-related thermoelectric effects in electronic transport through a nanoscopic system exhibiting magnetic anisotropy --with both  uniaxial and transverse components-- are studied theoretically in the linear response regime. In particular, a magnetic tunnel junction with a large-spin impurity --either a magnetic atom or a magnetic molecule-- embedded in the barrier is considered as an example. Owing to magnetic interaction with the impurity, conduction electrons traversing the junction can scatter on the impurity, which effectively can lead to angular momentum and energy exchange between the electrons and the impurity. As we show, such processes have a profound effect on the thermoelectric response of the system. Specifically, we present a detailed analysis of charge, spin and thermal conductance, together with the Seebeck and spin Seebeck coefficients (thermopowers).
Since the scattering mechanism also involves processes when electrons are inelastically scattered back to the same electrode, one can expect the flow of spin and energy also in the absence of charge transport through the junction. This, in turn, results in a finite spin thermopower, and the magnetic anisotropy plays a key role for this effect to occur.
\end{abstract}




\maketitle

\section{\label{Sec:Intro}Introduction}

One of the most promising routes towards maximizing the functional potential of nanoscopic electronic and spintronic devices relies on harnessing the interplay between transport of charge, spin and energy.~\cite{Dubi_Rev.Mod.Phys.83/2011,Goennenwein_NatureNanotechnol.7/2012,Bauer_NatureMater.11/2012,Maekawa_book_SC}
The interest in \emph{conventional} thermoelectric effects --those associated with transport of charge-- in nanoscopic systems was initiated more than two decades ago with experimental observations of such effects in mesoscopic conductors (like quantum wires~\cite{Gusev_JETPLett.51/1990,Gallagher_Phys.Rev.Lett.64/1990} and point contacts~\cite{Molenkamp_Phys.Rev.Lett.65/1990,Molenkamp_Phys.Rev.Lett.68/1992}),
and subsequently also in quantum dots.~\cite{Staring_Europhys.Lett.22/1993,Dzurak_SolidStateCommun.87/1993} On the other hand,
\emph{spin-dependent} thermoelectric effects --those connected with transport of spin due to a particle current-- have only recently become the subject of experiments, which so far have encompassed a wide range of nanosystems, including magnetic tunnel junctions,~\cite{LeBreton_Nature475/2011,Walter_NatureMater.10/2011,Liebing_Phys.Rev.Lett.107/2011,Lin_NatureCommun.3/2012} local~\cite{Dejene_Phys.Rev.B86/2012,Flipse_NatureNanotechnol.7/2012} and nonlocal~\cite{Bakker_Phys.Rev.Lett.105/2010,Erekhinsky_Appl.Phys.Lett.100/2012} spin valves, and others. Moreover, some new phenomena associated with electron spin and related to spin currents and/or spin accumulation, which can be considered as spin analogs of the corresponding conventional effects --so called \emph{spin} thermoelectric effects--  have also been studied.~\cite{Uchida_Nature455/2008,Bosu_Phys.Rev.B83/2011}
It is  worthy of note that spin currents can also be independently carried by magnon excitations,~\cite{Xiao_Phys.Rev.B81/2010,Adachi_Phys.Rev.B83/2011,Brechet_Phys.Rev.Lett.111/2013} so that corresponding spin thermoelectric effects can occur not only in metallic~\cite{Uchida_Nature455/2008,Bosu_Phys.Rev.B83/2011} or
semiconducting~\cite{Jaworski_NatureMater.9/2010} materials, but also in insulating magnets.~\cite{Uchida_NatureMater.9/2010,Uchida_Appl.Phys.Lett.97/2010}

The idea of spin-dependent thermoelectric effects was originally conceived by  Johnson and Silsbee.~\cite{Johnson_Phys.Rev.B35/1987} They predicted that in a tunnel junction with at least one of metallic electrodes being ferromagnetic, magnetization currents could be induced both electrically and thermally, and vice versa, thermal and electrical currents could also be induced magnetically. Later, similar concepts were considered theoretically  in a variety of nanosystems, including magnetic tunnel junctions,~\cite{Wang_Phys.Rev.B63/2001,McCann_Phys.Rev.B66/2002,Jansen_Phys.Rev.B85/2012,Misiorny_Phys.Rev.B89/2014} spin valves,~\cite{Hatami_Phys.Rev.B79/2009,Slachter_NaturePhys.6/2010,Yu_SolidStateCommun.150/2010} quantum dots,~\cite{Krawiec_Phys.Rev.B73/2006,Swirkowicz_Phys.Rev.B80/2009,Trocha_Phys.Rev.B85/2012,Weymann_Phys.Rev.B88/2013,
Lim_Phys.Rev.B88/2013,Haupt_PhysStatusSolidiB250/2013}, single-molecule-magnet junctions,~\cite{Wang_Phys.Rev.Lett.105/2010,Zhang_Appl.Phys.Lett.97/2010}
and multilayered systems.~\cite{Gravier_Phys.Rev.B73BR/2006,Gravier_Phys.Rev.B73/2006}

In the present paper we focus on a certain aspect of spin-related thermoelectric effects in nanoscopic systems that has not drawn much attention so far, namely on the influence of  magnetic anisotropy of a system on its thermoelectric properties. First, we consider the effect of magnetic anisotropy on the system's thermoelectric characteristics in the \emph{linear-response} regime. Second, we propose a scheme of how  magnetic anisotropy of a system  can be used to induce spin-thermal effects without actually transporting charge through the system. For this purpose, we consider a nanoscopic magnetic tunnel junction with a large-spin ($S>1/2$) impurity embedded in the barrier. In practice, such a model captures essential features of a simple planar magnetic tunnel junction, as well as of a setup involving the tip of a scanning tunneling microscope. In either case, the role of the impurity may be played, e.g., by a magnetic atom~\cite{Hirjibehedin_Science317/2007,Brune_Surf.Sci.603/2009} or a single-molecule magnet (SMM).~\cite{Kahle_NanoLett.12/2012}

\begin{figure}[t!!!]
   \includegraphics[width=\columnwidth]{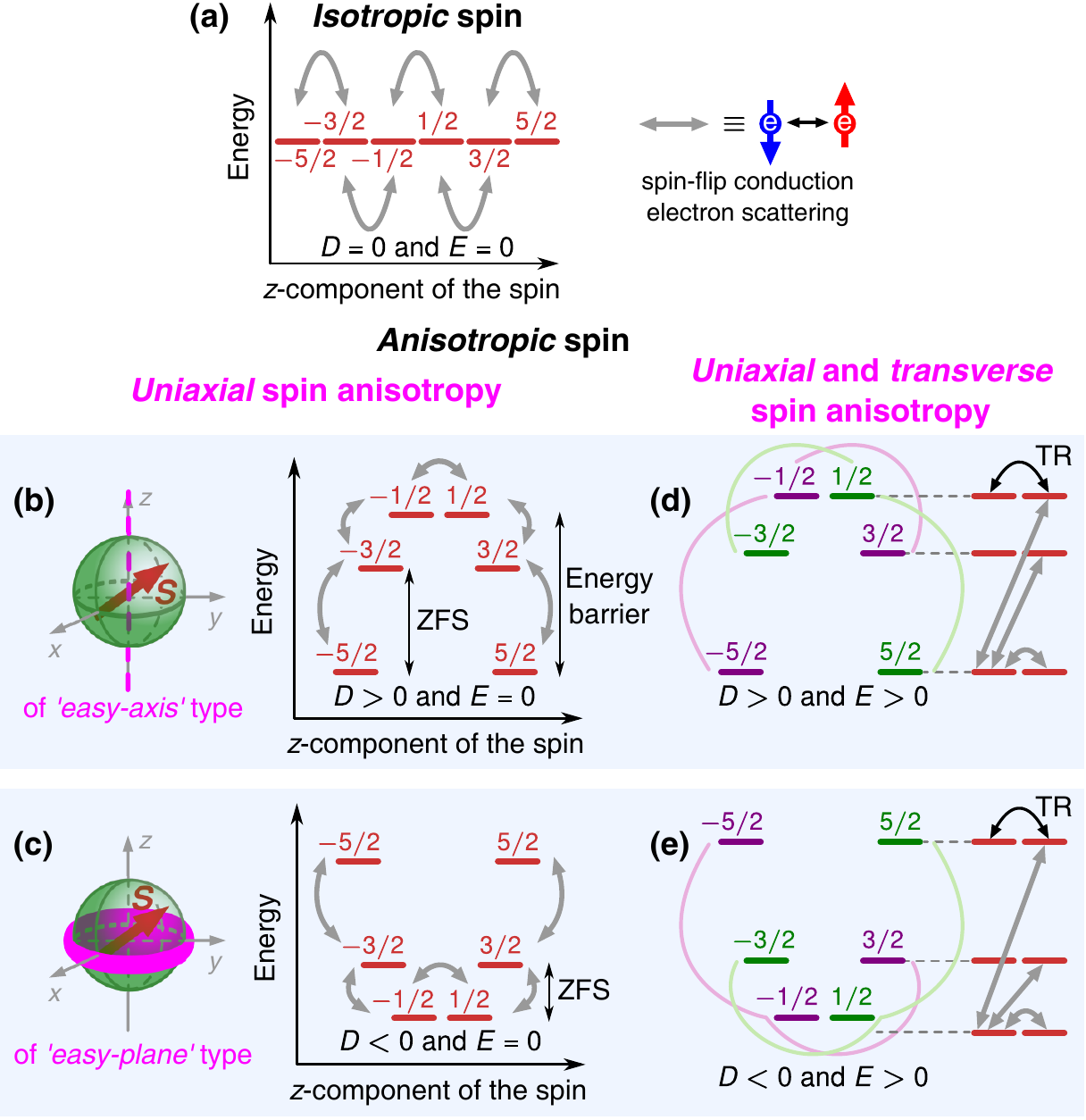}
    \caption{
    (color online)
    Influence of  magnetic anisotropy on the spectrum of an exemplary large-spin ($S=5/2$) impurity.
    (a) For an \emph{isotropic} spin, all $2S+1$ spin states are energetically degenerate. A conduction electron can flip its spin orientation when scattered on the impurity, which  results in transfer of the quantum of angular momentum $\hbar$ between the electron and impurity. The impurity undergoes then a transition between two neighboring spin states, $|\Delta S_z|=1$. These transitions are schematically marked by the double-sided gray arrows.
    However, spin-orbit  interaction in the presence of low symmetry of the impurity  usually leads to a \emph{spin-anisotropy}~(b)-(e). In the simplest case, the magnetic anisotropy  can be  \emph{uniaxial}~(b)-(c), which can be further distinguished into:
    (b) magnetic anisotropy of the `\emph{easy-axis}' type -- the spin prefers orientation along a specific axis (\emph{easy} axis, indicated by a vertical dashed bold line) without favoring any of the two orientations, which leads to an energy barrier for spin reversal;
    and
    (c) magnetic  anisotropy of the `\emph{easy-plane}' type -- the spin prefers orientation close to a specific plane
    (or in the plane for an integer $S$), often referred to as the \emph{easy} plane (indicated by a color plane perpendicular to the easy axis $z$).
     The uniaxial anisotropy  is very often also accompanied by a \emph{transverse} component (d)-(e), which introduces a mixing of the spin states.
     Due to such a mixing, depicted by thin color lines connecting states in (d)-(e), spin-flip scattering of conduction electrons can in principle transfer  the impurity between any two time-reversed     (TR)
      spin states. As an illustration,  possible transitions from one specific state are marked in (d)-(e) on the right side,
      with the states in the left column being time-reversed  with respect to the states in the right column.
        }
    \label{Fig:Fig1}
\end{figure}

Conduction electrons tunneling through the barrier can be scattered on the impurity, so their
energy  may be changed and spin orientation can be reversed. Such inelastic spin-flip scattering processes establish thus  a mechanism for energy and angular momentum transfer between the conduction electrons and the impurity, which, as we show below, is of key importance for
{\it spin} thermoelectric phenomena. There is, however, a fundamental difference between the behavior of \emph{spin-isotropic} and \emph{spin-anisotropic} impurities, as shown schematically in Fig.~\ref{Fig:Fig1}. In the former case, where the impurity spin does not  prefer any specific spatial orientation, all $2S+1$ spin states are energetically degenerate, so that only the spin angular momentum exchange can occur due to the electron scattering, see Fig.~\ref{Fig:Fig1}(a). On the other hand, in the spin-anisotropic case there are specific orientations of the impurity spin, which correspond to the lowest energy, as shown for instance in Fig.~\ref{Fig:Fig1}(b)-(e). Consequently, it is then possible that conduction electrons exchange not only  angular momentum with the impurity in a scattering process, but also energy can be transferred to/from the impurity during such an event. The main objective is to include these inelastic scattering processes in the description of thermoelectric phenomena. A similar problem has been analyzed in a recent paper,~\cite{Misiorny_Phys.Rev.B89/2014} but the considerations were limited to elastic scattering processes only. Thus, the corresponding description was applicable either to spin-isotropic impurities, or to spin-anisotropic case but in the low temperature limit, where only the two degenerate states of lowest energy could take part in transport. Here, we present a general description of the thermoelectricity, where the inelastic scattering processes are included in the linear response regime as well.

In Sec.~\ref{Sec:Overview_thermoelectrics} we present a brief overview of the conventional and spin thermoelectric effects. The system to be  considered, i.e., a magnetic tunnel junction with an impurity in the barrier is described in Sec.~\ref{Sec:Setup}. Transport kinetic coefficients are derived in Sec.~\ref{Sec:Transport}, while numerical results are presented and discussed in Sec.~\ref{Sec:Results}. Final conclusions are given in Sec.~\ref{Sec:Conclusions}.

\section{\label{Sec:Overview_thermoelectrics}Overview of conventional and spin-dependent thermoelectric effects}
In the situation when transport of charge, spin and energy occurs exclusively due to transfer of particles (e.g., electrons),
the physical origin of the thermoelectric effects under discussion relies on the
particle-hole asymmetry.~\cite{Barnard_book} When the particle-hole symmetry is present, a particle current in the presence of thermal gradient
is compensated by a hole current and the net charge current vanishes. This is not the case when the particle-hole symmetry is absent and the particle and hole currents do not compensate one another. As a result, there is a nonzero current associated with a thermal gradient. In order to ensure such a particle-hole asymmetry in planar junctions one needs to have an asymmetrical density of states (DOS) near the Fermi level in electrodes. In the case considered here, electrons traverse the barrier without entering the molecule, so this asymmetry is crucial. On the other hand, when conduction electrons tunnel through discrete levels of the  molecule (or a quantum dot), the DOS of the molecule is usually asymmetrical except some specific positions of the Fermi level, and this asymmetry is sufficient to generate thermoelectricity, also when DOS of the electrodes is symmetrical (and even independent of energy in the simplest case).

\begin{figure*}[t]
   \includegraphics[width=0.75\textwidth]{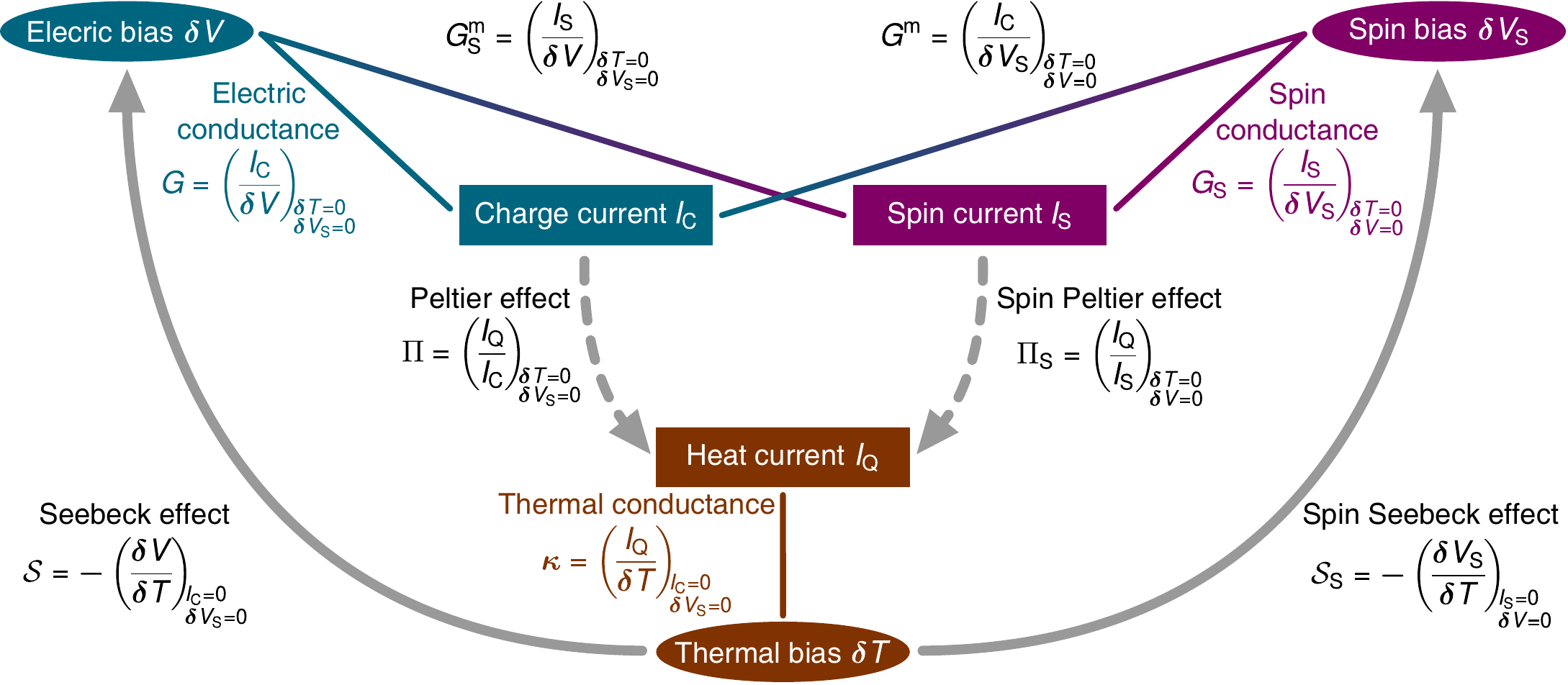}
    \caption{
    (color online)
    Systematic classification of the thermoelectric effects considered in the present work. Generally, in a biased system (with electric, spin, and thermal bias)
    a flow of  charge, spin and heat occurs,
    and the relation between a specific bias and currents is determined by the relevant \emph{conductances} (indicated by thin lines). These currents are not completely independent. For instance, under isothermal conditions, $\delta T=0$, a charge (spin) current is accompanied by a heat flow -- a phenomenon known  as the \emph{Peltier} and \emph{spin Peltier}  effects (dashed arrows), and characterized by the \emph{Peltier} and \emph{spin Peltier}  coefficients, $\Pi$ and $\Pi_\text{S}$, respectively. On the other hand, in the absence of a charge (spin) flow, $I_\text{C}=0$ ($I_\text{S}=0$),  a thermal bias can lead to charge (spin) accumulation, resulting in an electric and spin bias, respectively. This phenomenon is referred to as the \emph{Seebeck} and \emph{spin Seebeck}  effect (solid arrows), and is  quantified by the \emph{thermopower} $\mathcal{S}$ (or \emph{Seebeck} coefficient) and \emph{spin thermopower} $\mathcal{S}_\text{S}$ (or \emph{spin Seebeck} coefficient). We note that all the transport coefficients presented in the above figure are experimentally measurable.
    More detailed discussion can be found in Refs.~[\onlinecite{Goennenwein_NatureNanotechnol.7/2012}] and~[\onlinecite{Misiorny_Phys.Rev.B89/2014}].
    }
    \label{Fig:Fig2}
\end{figure*}

To make a brief survey of the conventional and spin-related thermoelectrics, let us consider a nanoscopic system attached to two ferromagnetic metallic electrodes. We will consider only electronic contributions to transport of charge, spin and energy. If the spin diffusion length  of conduction electrons is long enough relative to the system's length scale, such a system can usually be effectively represented by the transport model of two nonequivalent spin  channels. Generally, application of a constant voltage $\delta V$, spin voltage $\delta V_\text{S}$ and thermal bias $\delta T$ between the electrodes leads to  stationary charge transport of these three quantities is at the core for emergence of conventional and spin-related thermoelectric effects, as schematically depicted in  Fig.~\ref{Fig:Fig2}.

In the linear-response regime, transport through the two-terminal system under consideration can be formulated in terms of the \emph{kinetic coefficients} $\kcL_{kn}$,~\cite{Barnard_book,Mahan_book}
    \begin{equation}\label{Eq:L_matrix_def}
    \renewcommand{\arraystretch}{1.15}
        \left(\!\!
        \begin{array}{c}
        I_\text{C}
        \\
        I_\text{S}
        \\
        I_\text{Q}
        \end{array}
        \!\!
        \right)
    \!
    =
    \!
        \left(\!\!
        \begin{array}{ccc}
        e^2\kcL_{00} & e^2\kcL_{01} & e\kcL_{02}/T
        \\
        e\hbar\kcL_{10}/2 & e\hbar\kcL_{11}/2 & \hbar\kcL_{12}/(2T)
        \\
        e\kcL_{20} & e\kcL_{21} & \kcL_{22}/T
        \end{array}
        \!\!
        \right)
        \!\!\!
        \left(\!\!
        \begin{array}{c}
        \delta V
        \\
        \delta V_\text{S}
        \\
        \delta T
        \end{array}
        \!\!
        \right)
    ,
    \end{equation}
which satisfy the Onsager relation,~\cite{Onsager_Phys.Rev.37/1931_I,Onsager_Phys.Rev.38/1931_II,deGroot_book} $\kcL_{nk}=\kcL_{kn}$. Using the definitions shown in Fig.~\ref{Fig:Fig2}, one can  express the experimentally relevant parameters in terms of the above kinetic coefficients as follows.
\begin{enumerate}
\item[(i)]
Generalized electrical conductance matrix $\vec{G}$,~\cite{Swirkowicz_Phys.Rev.B80/2009}
\end{enumerate}
    \begin{equation}\label{Eq:G_def}
    \renewcommand{\arraystretch}{1.15}
    \vec{G}
    \equiv
        \left(\!\!
        \begin{array}{cc}
        G  & G^\text{m}
        \\
       G_\text{S}^\text{m} & G_\text{S}^{}
       \end{array}
        \!\!
        \right)
    =
        \left(\!\!
        \begin{array}{cc}
        e^2\kcL_{00}  & e^2\kcL_{01}
        \\
       e\hbar\kcL_{10}/2 & e\hbar\kcL_{11}/2
       \end{array}
        \!\!
        \right)
        ,
    \end{equation}
with $G^\text{m}$ and $G_\text{S}^\text{m}$ related due to the Onsager relation as $G_\text{S}^\text{m}=(\hbar/2e)G^\text{m}
=-(\hbar/2|e|)G^\text{m}$.
Because the electron charge is negative ($e<0$), one can immediately notice that $G^\text{m}$ and~$G_\text{S}^\text{m}$ must have opposite signs.
\begin{enumerate}
\item[(ii)]
Thermal conductance $\kappa$,
\end{enumerate}
    \begin{equation}\label{Eq:kappa_def}
    \kappa
    =
    \frac{1}{T}
    \Bigg[\kcL_{22}-\frac{\big(\kcL_{02}\big)^2}{\kcL_{00}}\Bigg]
    .
    \end{equation}
\begin{enumerate}
\item[(iii)]
Conventional $\Pi$ and spin  $\Pi_\text{S}$ Peltier coefficients,
\end{enumerate}
    \begin{equation}\label{Eq:Pi_def}
    \Pi=-\frac{1}{|e|}\frac{\kcL_{20}}{\kcL_{00}}
    \quad
    \textrm{and}
    \quad
    \Pi_\text{S}=\frac{2}{\hbar}\frac{\kcL_{21}}{\kcL_{11}}
    .
    \end{equation}
\begin{enumerate}
\item[(iv)]
Conventional $\mathcal{S}$ and spin $\mathcal{S}_\text{S}$ thermopowers, known also as the Seebeck and spin Seebeck coefficients,
\end{enumerate}
    \begin{equation}\label{Eq:S_def}
    \left\{
    \begin{aligned}
    \mathcal{S}
    =\ &
    \frac{\partial I_\text{C}/\partial\delta T}{G}
    =
    -\frac{1}{|e|T}\frac{\kcL_{02}}{\kcL_{00}}
    ,
    \\
    \mathcal{S}_\text{S}
    =\ &
    \frac{\partial I_\text{S}/\partial\delta T}{G_\text{S}}
    =
    -\frac{1}{|e|T}\frac{\kcL_{12}}{\kcL_{11}}
    .
    \end{aligned}
    \right.
    \end{equation}
Using the Onsager relation, one can easily note the relation between the Peltier and Seebeck coefficients, referred to as the \emph{Thompson's second relation},~\cite{deGroot_book}
    $
    \Pi=T\mathcal{S}
    $,
and its spin analog
    $
    \Pi_\text{S}=(2e/\hbar)T\mathcal{S}_\text{S}=-(2|e|/\hbar)/T\mathcal{S}_\text{S}$.
In addition, one can describe the overall (spin) thermoelectric efficiency of a system by means of the so-called (\emph{spin}) \emph{figure of merit}~$\ZT$~($\ZT_\text{S}$),
    \begin{equation}\label{Eq:ZT_def}
    \ZT=\frac{\mathcal{S}^2GT}{\kappa +\kappa_\text{ph}}
    \quad
    \textrm{and}
    \quad
    \ZT_\text{S}
    =
    \frac{2|e|}{\hbar}\frac{\mathcal{S}_\text{S}^2|G_\text{S}|T}{\kappa+\kappa_\text{ph}},
    \end{equation}
where $\kappa_\text{ph}$ is the phonon contribution to the heat conductance. This contribution,  however, will not be considered here.
In consequence, theoretical analysis of  the linear-response thermoelectric properties of a system can be presented in terms
 of the kinetic coefficients, derived in Sec.~\ref{Sec:Linear_response}.

\section{\label{Sec:Setup}Model system: magnetic tunnel junction with spin impurity}

\begin{figure}[t]
   \includegraphics[width=0.99\columnwidth]{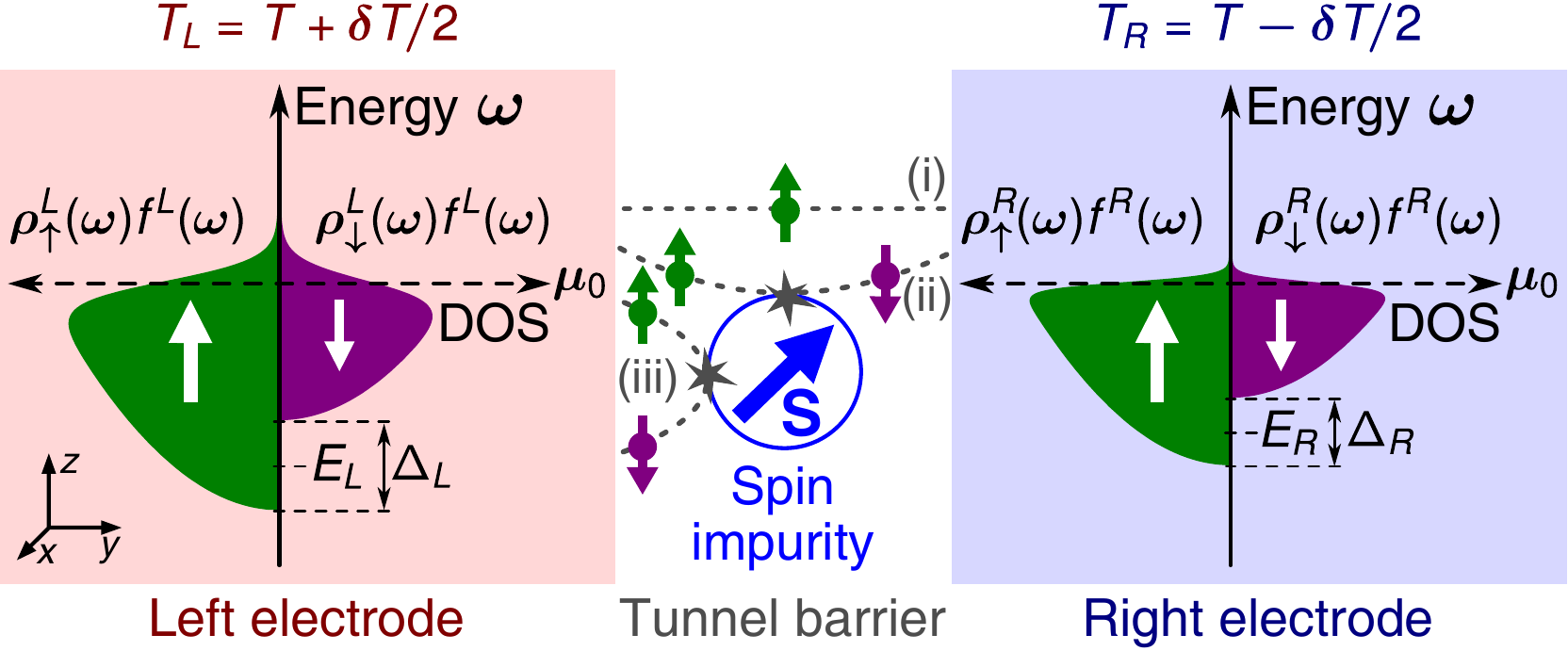}
    \caption{
    (color online)
    Schematic illustration of the model system under consideration: a magnetic tunnel junction with a large-spin impurity embedded in the tunnel barrier. Transport of electrons in such a junction can be driven by an external electric bias $\delta V$, spin bias $\delta V_\text{S}$, and the difference $\delta T$ between electrodes' temperatures $T_q$ [$q=L(\text{eft}),R(\text{ight})$].
    For conceptual simplicity we assume here  $\delta V=\delta V_\text{S}=0$.
    The  thermal bias $\delta T$ is included \emph{via} the Fermi-Dirac distribution functions $f^q(\omega)$, resulting in a  different smearing of densities of occupied states around the equilibrium electrochemical potential $\mu_0$ in each electrode.
    Apart from \emph{direct} tunneling between the electrodes [an example for spin-up electrons indicated by (i)], electrons can be also scattered by the impurity spin. As a result of such a process an electron can either tunnel to an opposite electrode (ii) or return to the same electrode (iii), and its spin can be conserved or  flipped [as shown for instance in (ii)-(iii)].
    }
    \label{Fig:Fig3}
\end{figure}

The model of a magnetic tunnel junction to be investigated here consists of two ferromagnetic metallic electrodes separated by an insulating  tunneling barrier, see Fig.~\ref{Fig:Fig3}. In general, transport of charge, spin and energy across such a junction, mediated  by tunneling of electrons, can occur either due to applied voltage (electric $\delta V$ or spin $\delta V_\text{S}$) or due to a thermal bias $\delta T$. Furthermore, if a spin impurity, represented by a spin operator $\hat{\vec{S}}=(\hat{S}_x,\hat{S}_y,\hat{S}_z)$, is incorporated into the barrier, apart from direct electron tunneling processes  between the electrodes, there are also processes where tunneling electrons are scattered on the impurity. As already mentioned in Sec.~\ref{Sec:Intro}, for a large-spin ($S>1/2$) and \emph{anisotropic} impurity,  this leads to transfer of angular momentum and energy between the impurity and scattered electrons, cf.~Fig.~\ref{Fig:Fig1}.

The  electrodes are modeled as reservoirs of itinerant, non-interacting electrons.
Hamiltonian of the electrodes takes the form
    $
     \hat{\Ham}_\text{el}=
     \sum_{qk\sigma}
     \varepsilon_{k\sigma}^q \hat{a}_{k\sigma}^{q\dag} \hat{a}_{k\sigma}^q
     $,
where $\hat{a}_{k\sigma}^{q\dag}$ ($\hat{a}_{k\sigma}^{q}$) is the electron creation (annihilation) operator and $\varepsilon_{k\sigma}^q$ is the conduction electron energy dispersion for the $q$th electrode [$q=L(\text{eft}),R(\text{ight})$], with $k$ and $\sigma$ denoting an orbital and electron spin index, respectively.
Moreover, since the transport effects to be discussed in the system under consideration  rely on the particle-hole asymmetry of DOSs around the equilibrium electrochemical potential $\mu_0$, we approximate the electrodes as Stoner magnets with parabolic dispersion relations,
so that both electrodes are characterized by a \emph{spin-dependent} DOS $\rho_\sigma^q(\omega)$,
    \begin{equation}\label{Eq:DOS}
    \rho_\sigma^q(\omega)
    =
    \Lambda_q
    \sqrt{\omega -\mu_0-E_q+\eta_\sigma\Delta_q/2}\geqslant0
    ,
    \end{equation}
for
$\omega\geqslant E_q +\mu_0 - \eta_\sigma\Delta_q/2$,
where $\sigma =+\,(-)$ refers to the majority (minority) electrons, $\eta_{+(-)}=\pm1$, $\Lambda_q$ is a material-dependent constant ($\sim0.1$ eV$^{-3/2}$), $E_q$ denotes the bottom edge of the conduction band in the $q$th electrode in the paramagnetic limit ($\Delta_q=0$), while $\Delta_q$ represents the  \emph{Stoner splitting}, i.e., the gap between  bottom edges of  the \emph{spin-majority} and \emph{spin-minority} conduction bands, see Fig.~\ref{Fig:Fig3}.
In addition, magnetic properties of the $q$th electrode can be conveniently  represented by \emph{spin polarization} at the Fermi level,~$P_q$, defined as
    \begin{equation}\label{}
    P_q
    =
    \frac{\rho_+^q(\omega =\mu_0)-\rho_-^q(\omega =\mu_0)}{\rho_+^q(\omega =\mu_0)+\rho_-^q(\omega =\mu_0)}
    .
    \end{equation}
Employing the above equation together with Eq.~(\ref{Eq:DOS}), one can easily find the relation between the spin polarization coefficient and the Stoner splitting,
    $
    \Delta_q
    =
    4P_q(\mu_0-E_q)/(1+P_q^2)
    $.
Therefore, in the following only $P_q$ will be used.
Moreover, two distinctive cases of the relative orientation of the electrodes' spin moments will be analyzed, namely the \emph{parallel} (shown in Fig.~\ref{Fig:Fig3}) and \emph{antiparallel} (with magnetic moment of the right electrode reversed) configuration.

Electron tunneling processes in the junction are  described by the Appelbaum Hamiltonian,~\cite{Appelbaum_Phys.Rev.Lett.17/1966,Appelbaum_Phys.Rev.154/1967}
    \begin{align}\label{Eq:Ham_tun}
    \hat{\Ham}_\text{tun}
    =\
    K
    \sum_{kk^\prime}
    \Big\{
    &
    \alpha_\text{d}
    \sum_{q\alpha}
    \hat{a}_{k\alpha}^{q\dag}
    \hat{a}_{k^\prime\alpha}^{\bar{q}}
    \nonumber\\
    +&
    \sum_{qq^\prime}
    \alpha_\text{ex}^{qq^\prime}
    \sum_{\alpha\beta}
    \hat{\vecG{\sigma}}_{\alpha\beta}\cdot\hat{\vec{S}}\:
    \hat{a}_{k\alpha}^{q\dag}
    \hat{a}_{k^\prime\beta}^{q^\prime}
    \Big\}
    ,
    \end{align}
where the notation  $\overline{q}$ should be understood as $\overline{L}\equiv R$ and $\overline{R}\equiv L$, whereas $\hat{\vecG{\sigma}}=(\hat{\sigma}_x,\hat{\sigma}_y,\hat{\sigma}_z)$ with  $\hat{\sigma}_i$ ($i=x,y,z$) denoting the Pauli matrices.
Thus, the first term of the Hamiltonian~(\ref{Eq:Ham_tun}) represents processes of direct tunneling across the junction. All other tunneling processes, where electrons interact magnetically with the impurity spin~$\hat{\vec{S}}$, either \emph{via} exchange coupling or direct dipolar interactions,~\cite{Hirjibehedin_Science317/2007} are included in the second term of $\hat{\Ham}_\text{tun}$.
Such processes  can be further divided into single- ($q=q^\prime$) and two-electrode ($q\neq q^\prime$)  ones.
Moreover, due to the interaction with the impurity, electron spin can flip ($\alpha\neq\beta$).
In the equation above, $\alpha_\text{ex}^{qq^\prime}\equiv\nu_q\nu_{q^\prime}\alpha_\text{ex}$, where a dimensionless parameter $\nu_q$ quantifies the coupling between the spin impurity and the $q$th electrode, while $K$ is the experimentally relevant energy parameter describing tunneling. We note that the ratio of dimensionless parameters $\alpha_\text{d}$ and $\alpha_\text{ex}$  gives the relative strength of the processes when electrons tunnel directly between electrodes with respect to those when they interact with the impurity.

The spin impurity in the barrier is described by the \emph{giant-spin} Hamiltonian,~\cite{Gatteschi_book}
    \begin{equation}\label{Eq:Ham_S}
    \hat{\Ham}_\text{imp}
    =
    -D\hat{S}_z^2+E\big(\hat{S}_x^2-\hat{S}_y^2\big)
    ,
    \end{equation}
with the first/second term describing the \emph{uniaxial}/\emph{transverse} magnetic anisotropy, and the parameters $D$ and $E$ being the relevant anisotropy constants.
Let us neglect for a moment the transverse term, $E=0$. The impurity Hamiltonian~(\ref{Eq:Ham_S}) is then diagonal in the basis of the eigenstates $\ket{S_z}$ of the spin operator $\hat{S}_z$. It can be easily seen that depending on the sign of the parameter $D$, the impurity spin  prefers the orientation either along the $z$-axis ($D>0$), see Fig.~\ref{Fig:Fig1}(b), or in the $xy$-plane ($D<0$), see Fig.~\ref{Fig:Fig1}(c). Especially interesting is the former case as it corresponds to formation of an energy barrier for spin reversal between two  metastable orientations (represented by the eigenstates $\ket{\pm S}$) along the $z$-axis, referred then to  as the system's \emph{easy axis}.
Importantly, in order to surmount the barrier, the impurity spin has to undergo a series of transitions \emph{via} all consecutive states $\ket{S_z}$ separating the metastable states $\ket{\pm S}$.
On the other hand, once $E\neq0$, the simple picture introduced above doesn't hold as the transverse term allows for mixing of states $\ket{S_z}$, which is schematically illustrated in Fig.~\ref{Fig:Fig1}(d)-(e). In particular, in the present situation each of the $2S+1$ eigenstates~$\ket{\chi}$ of the impurity Hamiltonian~(\ref{Eq:Ham_S}), $\hat{\Ham}_\text{imp}\ket{\chi}=\varepsilon_\chi\ket{\chi}$, is a linear combination of states~$\ket{S_z}$. As a result, direct transitions between different spin states on the opposite sides of the spin-reversal barrier become in principle permitted.

Finally, we assume $E>0$, so that  the orientation of the impurity spin along the $y$-axis is energetically more favored than along the $x$-axis. Apart from this, we assume that  the two magnetic anisotropy constants are customarily related as $0\leqslant E/|D|\leqslant1/3$. This condition  means that the $z$-axis is always associated with the dominating (uniaxial) component of magnetic anisotropy. We also note that since our main objective is to present the general concept of how the magnetic anisotropy can modify the thermoelectric properties of a nanosystem, we restrict our considerations to the situation when the impurity's easy axis is collinear with the electrodes' spin moments.

\section{\label{Sec:Transport}Transport characteristics}

\subsection{General formulation}
Since the effective coupling of the impurity to external electrodes is assumed to be weak, transport characteristics of the system can be derived in terms of the approach based on the relevant master equation. This approach allows for calculating balanced flows of charge, spin and energy between the electrodes due to   tunneling electrons ($e<0$), with the relevant currents defined as follows:~\cite{Misiorny_Phys.Rev.B89/2014}
\begin{enumerate}
\item[(i)]
charge current
\end{enumerate}
    \begin{multline}\label{Eq:I_C_def}
    I_\text{C}
    =
    e\Gamma
    \!\!
    \sum_{q\sigma\chi\chi^\prime}
    \!\!
    \eta_q
    \Big[
    \delta_{\chi\chi^\prime}
    \mathcal{W}_{\sigma,\chi}^\text{sc}
    \aT_{\overline{q}\sigma,\chi}^{(0)q\sigma,\chi}
    \\
    +
    \mathcal{W}_{\sigma,\chi^\prime\chi}^\text{sf}
    \aT_{\overline{q}\overline{\sigma},\chi^\prime}^{(0)q\sigma,\chi}
    \Big],
    \end{multline}
\begin{enumerate}
\item[(ii)]
spin current
\end{enumerate}
    \begin{multline}\label{Eq:I_S_def}
    I_\text{S}
    =
    \frac{\hbar\Gamma}{2}
    \!\!
    \sum_{q\sigma\chi\chi^\prime}
    \!\!
    \eta_\sigma
    \eta_q
    \Big[
    \delta_{\chi\chi^\prime}
    \mathcal{W}_{\sigma,\chi}^\text{sc}
    \aT_{\overline{q}\sigma,\chi}^{(0)q\sigma,\chi}
    \\
    +
    \mathcal{V}_{q\sigma,\chi^\prime\chi}^\text{sf}
    \aT_{q\overline{\sigma},\chi^\prime}^{(0)q\sigma,\chi}
    \Big]
    ,
    \end{multline}
\begin{enumerate}
\item[(iii)]
heat current
\end{enumerate}
    \begin{align}\label{Eq:I_Q_def}
    \hspace*{-4pt}
    I_\text{Q}
    =
    \Gamma\!\!
    \sum_{q\sigma\chi\chi^\prime}
    \!\!
    \eta_q
    \Big[
    \delta_{\chi\chi^\prime}
    \mathcal{W}_{\sigma,\chi}^\text{sc}
    \aT_{\overline{q}\sigma,\chi}^{(1)q\sigma,\chi}
    +
    \mathcal{W}_{\sigma,\chi^\prime\chi}^\text{sf}
    \aT_{\overline{q}\overline{\sigma},\chi^\prime}^{(1)q\sigma,\chi}
    &
    \nonumber\\
    -
    \mathcal{V}_{q\sigma,\chi^\prime\chi}^\text{sf}
    \Big(
    \case{1}{2}\Delta_{\chi\chi^\prime}+\case{1}{2}\big(\mu_\sigma^q-\mu_{\overline{\sigma}}^q\big)
    \!
    \Big)
    \aT_{q\overline{\sigma},\chi^\prime}^{(0)q\sigma,\chi}
    &
    \Big]
    .
    \end{align}
Here, $\Gamma\equiv\pi K^2/\hbar$, $\Delta_{\chi\chi^\prime}=\varepsilon_{\chi}-\varepsilon_{\chi^\prime}$, and $\mu_\sigma^q=\mu_0+e\eta_q(\delta V+\eta_\sigma \delta V_\text{S})/2$ stands for  the spin-dependent  electrochemical potential of the $q$th electrode,
with $\eta_{L(R)}\equiv\pm1$ and $\eta_{\uparrow(\downarrow)}=\pm1$,
together with $\mu_0=(\mu_\sigma^L+\mu_\sigma^R)/2$ denoting the electrochemical potential at equilibrium and
$\delta V$ ($\delta V_\text{S}$) representing the electric (spin) voltage bias.
Furthermore, $\aT_{q^\prime\sigma^\prime,\chi^\prime}^{(n)q\sigma,\chi}$ is a function describing the transfer of an electron of initial spin $\sigma$ from the $q$th electrode into  the $q^\prime$th electrode which the electron enters with spin~$\sigma^\prime$. The possible flip of the electron's spin orientation ($\sigma\rightarrow\sigma^\prime=\overline{\sigma}$) appears due to electron scattering on the impurity, associated with transition of the impurity magnetic state from $\ket{\chi}$ to~$\ket{\chi^\prime}$. This function is given by
    \begin{equation}\label{Eq:funT_def}
    \aT_{q^\prime\sigma^\prime,\chi^\prime}^{(n)q\sigma,\chi}
    =
    \prob_\chi
    \Phi_{q^\prime\!\sigma^\prime}^{(n)q\sigma}\!\big(\Delta_{\chi\chi^\prime}\!\big)
    -
    \prob_{\chi^\prime}
    \Phi_{q\sigma}^{(n)q^\prime\! \sigma^\prime}\!\!\big(\Delta_{\chi^\prime\chi}\big)
    ,
    \end{equation}
with $\prob_\chi$ being the probability of finding the spin impurity in the state $\ket{\chi}$, and
    \begin{multline}\label{Eq:Phi_fun}
    \Phi_{q^\prime\!\sigma^\prime}^{(n)q\sigma}\!\big(\Delta_{\chi\chi^\prime}\!\big)
    =
    \!\!
    \int\!\!\text{d}\omega
    \,
    \rho_\sigma^q\big(\omega\big)
    \rho_{\sigma^\prime}^{q^\prime}\big(\omega+\Delta_{\chi\chi^\prime}\!\big)
    \\
    \times
    \Big[
    \omega
    +\case{1}{2}\Delta_{\chi\chi^\prime}
    -\case{1}{2}\big(\mu_\sigma^q+\mu_{\sigma^\prime}^{q^\prime}\big)
    \Big]^{\!n}
    \\
    \times
    f_{\sigma}^q\big(\omega\big)
    \!
    \Big[1-f_{\sigma^\prime}^{q^\prime}\!\big(\omega+\Delta_{\chi\chi^\prime}\!\big)\Big]
    .
    \end{multline}
In the equation above,
    $
    f_{\sigma}^q\big(\omega\big)
    =
    \big\{1+\textrm{exp}\big[(\omega-\mu_\sigma^q)/T_q\big]\big\}^{-1}
    $
is the spin-dependent Fermi-Dirac distribution function for the $q$th electrode, with $T_q=T+\eta_q\delta T$ denoting the temperature of the electrode
expressed in energy units (i.e.,  $k_\text{B}\equiv 1$).
Though in the following we express the temperature in the energy units, the original units will be  restored when presenting numerical results.
Moreover, the matrix element
    \begin{equation}\label{Eq:Wsc_def}
    \mathcal{W}_{\sigma,\chi}^\text{sc}
    =
    \alpha_\text{d}^2
    +
    \big(\alpha_\text{ex}^{LR}\big)^{\!2}
    \big|\mathbb{S}_{\chi\chi}^z\big|^2
    +
    2
    \eta_\sigma
    \alpha_\text{d}\alpha_\text{ex}^{LR}
    \mathbb{S}_{\chi\chi}^z
    ,
    \end{equation}
where $\mathbb{S}_{\chi^\prime\chi}^k\equiv\bra{\chi^\prime}\hat{S}_k\ket{\chi}$ for $k=z,\pm$,
quantifies the effect of electron tunneling processes between the left and right electrode without spin reversal.
The first term of Eq.~(\ref{Eq:Wsc_def}) accounts for the \emph{direct} tunneling processes, whereas the second term corresponds to the \emph{indirect}  \emph{spin-conserving} scattering on the impurity.
Moreover, one could also expect the interference between the two aforementioned tunneling paths, which is described by the last term of Eq.~(\ref{Eq:Wsc_def}). However, under the linear-response conditions this term will actually play no role, as it will turn out later.
The  \emph{spin-flip} tunneling processes are, in turn, described by the matrix elements
    \begin{equation}\label{Eq:Wsf_Vsf_def}
    \renewcommand{\arraystretch}{1.15}
    \left(\!\!
    \begin{array}{c}
    \mathcal{W}_{\sigma,\chi^\prime\chi}^\text{sf}
    \\
    \mathcal{V}_{q\sigma,\chi^\prime\chi}^\text{sf}
    \end{array}
    \!\!
    \right)
    =
    \left(
    \!\!
    \begin{array}{*{20}{c}}
    \big(\alpha_\text{ex}^{LR}\big)^{\!2}
    \\
     \big(\alpha_\text{ex}^{qq}\big)^{\!2}
    \end{array}
    \!\!
    \right)
    \Big[
    \delta_{\sigma\downarrow}\big|\mathbb{S}_{\chi^\prime\chi}^-\big|^2
    +
    \delta_{\sigma\uparrow}\big|\mathbb{S}_{\chi^\prime\chi}^+\big|^2
    \Big]
    .
    \end{equation}
For the sake of notational clarity, in Eq.~(\ref{Eq:Wsf_Vsf_def}) we have further distinguished the two- ($\aW^\text{sf}$) and single-electrode ($\aV^\text{sf}$)  electron tunneling processes. Importantly, it is worth emphasizing that in the latter case  spin-flip processes can contribute only to transport of spin and energy across the junction, cf. Eqs.~(\ref{Eq:I_C_def})-(\ref{Eq:I_Q_def}).

Finally, as one can notice from Eq.~(\ref{Eq:funT_def}), the usage of the expressions~\mbox{(\ref{Eq:I_C_def})-(\ref{Eq:I_Q_def})} requires the knowledge of the stationary probabilities~$\prob_{\chi}$. At equilibrium, these probabilities are given by the Gibbs distribution,
    \begin{equation}
    \prob_\chi\big|_\text{eq}
    \equiv
    \widetilde{\prob}_\chi
    =
    \frac{1}{\mathcal{Z}}
    \,
    \text{exp}\big[-\varepsilon_\chi/T\big]
    ,
    \end{equation}
with the partition function
    $
    \mathcal{Z}
    =
    \sum_\chi
    \text{exp}\big[-\varepsilon_\chi/T\big].
    $
Note the notation $\bullet\big|_\text{eq}\equiv\widetilde{\bullet}$ we will use henceforth interchangeably for
denoting the quantity at equilibrium, that is  at~$\delta V=\delta V_\text{S}=\delta T=0$.
On the other hand, out of equilibrium the probabilities $\prob_\chi$ can in general be found as a solution to the set of stationary master equations
    \begin{equation}\label{eq:Master_eq}
    \frac{\text{d}\prob_\chi}{\text{d}t}=0=
    \sum_{\chi^\prime}
    \Big[
    \gamma_{\chi^\prime\chi}\prob_{\chi^\prime}
    -
    \gamma_{\chi\chi^\prime}\prob_{\chi}
    \Big]
    ,
    \end{equation}
with the probability normalization condition $\sum_\chi\prob_\chi=1$, and
    \begin{align}
    \gamma_{\chi\chi^\prime}=
    2\Gamma
    \sum_{q\sigma}
    \Big[
    &
    \mathcal{W}_{\sigma,\chi^\prime\chi}^\text{sf}
    \Phi_{\overline{q}\overline{\sigma}}^{(0)q\sigma}\!\big(\Delta_{\chi\chi^\prime}\!\big)
    \nonumber\\
    +\ &
    \mathcal{V}_{q\sigma,\chi^\prime\chi}^\text{sf}
    \Phi_{q\overline{\sigma}}^{(0)q\sigma}\!\big(\Delta_{\chi\chi^\prime}\!\big)
    \Big]
    .
    \end{align}
%

\subsection{Linear response regime: kinetic coefficients\label{Sec:Linear_response}}
The kinetic coefficients $\kcL_{kn}$, cf.~Eq.~(\ref{Eq:L_matrix_def}), can be derived by linearization of the Eqs.~(\ref{Eq:I_C_def})-(\ref{Eq:I_Q_def}), which we precede with the substitution
    \begin{equation}\label{eq:P_LinExpand}
    \prob_\chi
    =
    \widetilde{\prob}_\chi
    \Big[
    1
    +
    \Psi_{x_n}^\chi\,
    \frac{
    e^{\delta_{n0}+\delta_{n1}}
    }{
    T
    }
    x_n
    \Big]
    .
    \end{equation}
Here, the shorthand notation $x_0\equiv\delta V$, $x_1\equiv\delta V_\text{S}$ and $x_2\equiv\delta T$ has been introduced. Employing the identity
    $
    \widetilde{\prob}_{\chi^\prime}
    =
    \widetilde{\prob}_{\chi}\,
    \text{exp}\big[\Delta_{\chi\chi^\prime}/T\big]
    $
together with the symmetry property
    \begin{equation}
    \widetilde{\Phi}_{q^\prime\overline{\sigma}}^{(n)q\sigma}\!\big(\Delta_{\chi\chi^\prime}\!\big)
    =
    \widetilde{\Phi}_{q\sigma}^{(n)q^\prime \overline{\sigma}}\!\big(\!-\Delta_{\chi\chi^\prime}\!\big)
    \,
    \text{exp}\big[\Delta_{\chi\chi^\prime}/T\big]
    ,
    \end{equation}
we obtain
\begin{widetext}
    \begin{align}\label{Eq:L_0n_def}
    \kcL_{0n}
    =\ &
    \frac{T^{\delta_{n2}}}{e^{\delta_{n0}+\delta_{n1}}}
    \Gamma
    \!
    \sum_{q\sigma\chi\chi^\prime}\!\!
    \eta_q
    \widetilde{\prob}_\chi
    \Bigg[
    2
    \delta_{\chi\chi^\prime}
    \mathcal{W}_{\sigma,\chi}^\text{sc}
    \frac{
    \partial \Phi_{{\overline{q}}\sigma}^{(0)q\sigma}\!\big(0\big)
    }{
    \partial x_n
    }\Big|_\text{eq}
    \nonumber\\
    &\hspace*{100pt}
    +
    \!
    \mathcal{W}_{\sigma,\chi^\prime\chi}^\text{sf}
    \Bigg\{
    \frac{
    \partial\Upsilon_{{\overline{q}}\overline{\sigma}}^{(0)q\sigma}\!\big(\Delta_{\chi\chi^\prime}\!\big)
    }{
    \partial x_n
    }\Big|_\text{eq}
    \!
    +
    \frac{e^{\delta_{n0}+\delta_{n1}}}{T}
    \Big[
    \Psi_{x_n}^\chi
    -
    \Psi_{x_n}^{\chi^\prime}
    \Big]
    \widetilde{\Phi}_{{\overline{q}}\overline{\sigma}}^{(0)q\sigma}\!\big(\Delta_{\chi\chi^\prime}\!\big)
    \!
    \Bigg\}
    \Bigg],
    \\
    \label{Eq:L_1n_def}
    \kcL_{1n}
    =\ &
    \frac{T^{\delta_{n2}}}{e^{\delta_{n0}+\delta_{n1}}}
    \Gamma
   \!\!
    \sum_{q\sigma\chi\chi^\prime}\!\!
    \eta_\sigma
    \eta_q
    \widetilde{\prob}_\chi
    \Bigg[
    2
    \delta_{\chi\chi^\prime}
    \mathcal{W}_{\sigma,\chi}^\text{sc}
    \frac{
    \partial \Phi_{{\overline{q}}\sigma}^{(0)q\sigma}\!\big(0\big)
    }{
    \partial x_n
    }\Big|_\text{eq}
    \nonumber\\
    &\hspace*{100pt}
    +
    \mathcal{V}_{q\sigma,\chi^\prime\chi}^\text{sf}
    \Bigg\{
    \frac{
    \partial\Upsilon_{q\overline{\sigma}}^{(0)q\sigma}\!\big(\Delta_{\chi\chi^\prime}\!\big)
    }{
    \partial x_n
    }\Big|_\text{eq}
    +
    \frac{e^{\delta_{n0}+\delta_{n1}}}{T}
    \Big[
    \Psi_{x_n}^\chi
    -
    \Psi_{x_n}^{\chi^\prime}
    \Big]
    \widetilde{\Phi}_{q\overline{\sigma}}^{(0)q\sigma}\!\big(\Delta_{\chi\chi^\prime}\!\big)
    \Bigg\}\!
    \Bigg],
    \\
    \label{Eq:L_2n_def}
    \kcL_{2n}
    =\ &
    \frac{T^{\delta_{n2}}}{e^{\delta_{n0}+\delta_{n1}}}
    \Gamma
    \!\!
    \sum_{q\sigma\chi\chi^\prime}\!\!
    \eta_q
    \widetilde{\prob}_\chi
    \Bigg[
    2
    \delta_{\chi\chi^\prime}
    \mathcal{W}_{\sigma,\chi}^\text{sc}
    \frac{
    \partial\Phi_{\overline{q}\sigma}^{(1)q\sigma}\!\big(0\big)
    }{
    \partial x_n
    }\Big|_\text{eq}
                    \nonumber\\
    &\hspace*{100pt}
    +
    \mathcal{W}_{\sigma,\chi^\prime\chi}^\text{sf}
    \Bigg\{
    \frac{
    \partial\Upsilon_{\overline{q}\overline{\sigma}}^{(1)q\sigma}\!\big(\Delta_{\chi\chi^\prime}\!\big)
    }{
    \partial x_n
    }\Big|_\text{eq}
    +
    \frac{e^{\delta_{n0}+\delta_{n1}}}{T}
    \Big[
    \Psi_{x_n}^\chi
    -
    \Psi_{x_n}^{\chi^\prime}
    \Big]
    \widetilde{\Phi}_{\overline{q}\overline{\sigma}}^{(1)q\sigma}\!\big(\Delta_{\chi\chi^\prime}\!\big)
    \Bigg\}
                     \nonumber\\
    &\hspace*{100pt}
    -
    \frac{\Delta_{\chi\chi^\prime}}{2}
    \mathcal{V}_{q\sigma,\chi^\prime\chi}^\text{sf}
    \Bigg\{
    \frac{
    \partial\Upsilon_{q\overline{\sigma}}^{(0)q\sigma}\!\big(\Delta_{\chi\chi^\prime}\!\big)
    }{
    \partial x_n
    }\Big|_\text{eq}
    \!\!
    +
    \frac{e^{\delta_{n0}+\delta_{n1}}}{T}
    \Big[
    \Psi_{x_n}^\chi
    -
    \Psi_{x_n}^{\chi^\prime}
    \Big]
    \widetilde{\Phi}_{q\overline{\sigma}}^{(0)q\sigma}\!\big(\Delta_{\chi\chi^\prime}\!\big)
    \Bigg\}\!
    \Bigg]
    .
    \end{align}
\end{widetext}
In order to keep a compact form of the above expressions for the kinetic coefficients,  we have introduced the auxiliary function
    \begin{multline}
    \Upsilon_{q^\prime\overline{\sigma}}^{(n)q\sigma}\!\big(\Delta_{\chi\chi^\prime}\!\big)
    =
    \Phi_{q^\prime \overline{\sigma}}^{(n)q\sigma}\!\big(\Delta_{\chi\chi^\prime}\!\big)
    \\
    -
    \Phi_{q\sigma}^{(n)q^\prime \overline{\sigma}}\!\big(-\Delta_{\chi\chi^\prime}\!\big)
    \exp\big[\Delta_{\chi\chi^\prime}/T\big]
    .
    \end{multline}
As one can see, Eqs.~(\ref{Eq:L_0n_def})-(\ref{Eq:L_2n_def})
involve terms that are proportional to $\Psi_{x_n}^\chi -\Psi_{x_n}^{\chi^\prime}$, that is to the difference between linear terms in the Taylor expansion of the probabilities of two different impurity states $\ket{\chi}$ and $\ket{\chi^\prime}$.
Finding these directly from Eq.~(\ref{eq:Master_eq}) generally proves to be cumbersome (especially if the impurity spin $S$ is large). However, the master equation~(\ref{eq:Master_eq}) in the stationary limit is equivalent to the set of detailed balance equations (one for each state~$\ket{\chi^\prime}$):~\cite{Beenakker_Phys.Rev.B44/1991}
    $
    \gamma_{\chi\chi^\prime}\mathcal{P}_{\chi}
    =
    \gamma_{\chi^\prime\chi}\mathcal{P}_{\chi^\prime}
    $.
Linearization of this equation, after substitution of Eq.~(\ref{eq:P_LinExpand}) and then application of the identities already used to derive the expressions for kinetic coefficients~(\ref{Eq:L_0n_def})-(\ref{Eq:L_2n_def}),
yields
    \begin{align}
    \Psi_{x_n}^\chi
    -
    \Psi_{x_n}^{\chi^\prime}
    =\ &
    -
    \frac{
    T
    \Omega_{\chi^\prime\chi}
    }{
    e^{\delta_{n0}+\delta_{n1}}
    }
    \sum_{q\sigma}
    \Bigg[
    \mathcal{W}_{\sigma,\chi^\prime\chi}^\text{sf}
    \frac{\partial \Upsilon_{\overline{q}\overline{\sigma}}^{(0)q\sigma}\!\big(\Delta_{\chi\chi^\prime}\!\big)}{\partial x_n}\Big|_\text{eq}
    \nonumber\\
    &\hspace*{30pt}
    +
    \mathcal{V}_{q\sigma,\chi^\prime\chi}^\text{sf}
    \frac{\partial\Upsilon_{q\overline{\sigma}}^{(0)q\sigma}\!\big(\Delta_{\chi\chi^\prime}\!\big)}{\partial x_n}\Big|_\text{eq}
    \Bigg]
    ,
    \end{align}
where
    \begin{equation}\label{Eq:Omega}
    \Omega_{\chi^\prime\chi}
    =
    \Big\{
     \sum_{q\sigma}
    \Big[
    \aWW_{q\sigma,\chi^\prime\chi}^{(0)\text{sf}}
    +
    \aVV_{q\sigma,\chi^\prime\chi}^{(0)\text{sf}}
    \Big]
    \Big\}^{-1}
    ,
    \end{equation}
and
    \begin{equation}\label{Eq:notation_WV}
    \renewcommand{\arraystretch}{1.5}
    \left\{
    \begin{array}{l}
    \aWW_{q\sigma,\chi^\prime\chi}^{(n)\text{sc}}
    =
    \delta_{\chi\chi^\prime}
     \mathcal{W}_{\sigma,\chi}^\text{sc}
    \widetilde{\Phi}_{\overline{q}\sigma}^{(n)q\sigma}\!\big(0\big)
    ,
    \\
    \aWW_{q\sigma,\chi^\prime\chi}^{(n)\text{sf}}
    =
    \mathcal{W}_{\sigma,\chi^\prime\chi}^\text{sf}
    \widetilde{\Phi}_{\overline{q}\overline{\sigma}}^{(n)q\sigma}\!\big(\Delta_{\chi\chi^\prime}\!\big)
    ,
    \\
    \aVV_{q\sigma,\chi^\prime\chi}^{(n)\text{sf}}
    =
    \mathcal{V}_{q\sigma,\chi^\prime\chi}^\text{sf}
    \widetilde{\Phi}_{q\overline{\sigma}}^{(n)q\sigma}\!\big(\Delta_{\chi\chi^\prime}\!\big)
    .
    \end{array}
    \right.
    \end{equation}
Note the new compact notation for the matrix elements $\aWW_{q\sigma,\chi^\prime\chi}^{(n)\text{sf}}$ and $\aVV_{q\sigma,\chi^\prime\chi}^{(n)\text{sf}}$
 introduced above. Consequently, it can be noticed that in the final step, before the explicit expression for $\kcL_{kn}$ can be written down, one eventually has to find derivatives of the form
    $
    \big(
    \partial
    \Phi_{q^\prime \overline{\sigma}}^{(n)q\sigma}\!\big(\Delta_{\chi\chi^\prime}\!\big)
    /
    \partial
    x_n
    \big)|_\text{eq}
    $.
Interestingly, we find that all the derivatives in question can  be  actually expressed in terms of the $\widetilde{\Phi}$-functions, and thus the notation~(\ref{Eq:notation_WV}) will prove especially helpful. In particular, using that
    \begin{multline}\label{Eq:FD_fun_deriv}
    \hspace*{-1pt}
    \frac{\partial f_{\sigma}^q\big(\omega+\Delta\big)}{\partial x_n}\Big|_\text{eq}
    \!\!
    =
    \eta_q
    \eta_\sigma^{\delta_{n1}}
    \frac{e^{\delta_{n0}+\delta_{n1}}}{2T^{1+\delta_{n2}}}
    \big(\omega+\Delta-\mu_0\big)^{\!\delta_{n2}}
    \\
    \times
    f\big(\omega+\Delta\big)
    \Big[
    1-
    f\big(\omega+\Delta\big)
    \Big]
    ,
    \end{multline}
with $f(\omega)\equiv f_\sigma^q(\omega)\big|_\text{eq}$,
we obtain
    \begin{equation}
    \frac{\partial \Phi_{\overline{q}\sigma}^{(k)q\sigma}\!\big(0\big)}{\partial x_n}\Big|_\text{eq}
    =
    \eta_q
    \eta_\sigma^{\delta_{n1}}
    \frac{e^{\delta_{n0}+\delta_{n1}}}{2T^{1+\delta_{n2}}}
    \widetilde{\Phi}_{\overline{q}\sigma}^{(k+\delta_{n2})q\sigma}\!\big(0\big)
    ,
    \end{equation}
and after some more laborious, though straightforward, transformations we arrive at
    \begin{multline}
    \hspace*{-10pt}
    \frac{
    \partial\Upsilon_{q^\prime\overline{\sigma}}^{(0)q\sigma}\!\big(\Delta\big)
    }{
    \partial x_n
    }\Big|_\text{eq}
    =
    \Big[
    \big(
    \delta_{n0}+\delta_{n2}
    \big)
    \delta_{q^\prime\overline{q}}
    +
    \delta_{n1}
    \delta_{q^\prime\! q}\,
    \eta_\sigma
    \Big]
    \\
    \hspace*{-8pt}
    \times
    \eta_q
    \frac{e^{\delta_{n0}+\delta_{n1}}}{T^{1+\delta_{n2}}}
    \widetilde{\Phi}_{q^\prime\overline{\sigma}}^{(\delta_{n2})q\sigma}\!\big(\Delta\big)
    -
    \delta_{n2}
    \delta_{q^\prime\! q}\,
    \eta_q
     \frac{\Delta}{2T^2}
    \widetilde{\Phi}_{q\overline{\sigma}}^{(0)q\sigma}\!\big(\Delta\big)
    ,
    \hspace*{-8pt}
    \end{multline}
and
    \begin{equation}
    \hspace*{-5pt}
    \frac{
    \partial\Upsilon_{\overline{q}\overline{\sigma}}^{(1)q\sigma}\!\big(\Delta\big)
    }{
    \partial x_n
    }\Big|_\text{eq}
    =
    \big(
    \delta_{n0}+\delta_{n2}
    \big)
    \eta_q
    \frac{e^{\delta_{n0}}}{T^{1+\delta_{n2}}}
    \widetilde{\Phi}_{\overline{q}\overline{\sigma}}^{(1+\delta_{n2})q\sigma}\!\big(\Delta\big)
    .
    \hspace*{-3pt}
    \end{equation}
Finally, combining all the above expressions, the kinetic coefficients take the form
\begin{widetext}
\vspace*{-15pt}
    \begin{align}\label{Eq:L00}
    \kcL_{00}
    =\ &
    \frac{\Gamma}{T}
    \sum_{\chi\chi^\prime}
    \widetilde{\prob}_\chi
    \Bigg[
    \sum_{q\sigma}
    \Big(
    \aWW_{q\sigma,\chi^\prime\chi}^{(0)\text{sc}}
    +
    \aWW_{q\sigma,\chi^\prime\chi}^{(0)\text{sf}}
    \Big)
    -
    \Omega_{\chi^\prime\chi}
    \Big(
    \sum_{q\sigma}
    \eta_q
    \aWW_{q\sigma,\chi^\prime\chi}^{(0)\text{sf}}
    \Big)^{\!2}
    \Bigg]
    ,
    \\
    \label{Eq:L11}
    \kcL_{11}
    =\ &
    \frac{\Gamma}{T}
    \sum_{\chi\chi^\prime}
    \widetilde{\prob}_\chi
    \Bigg[
    \sum_{q\sigma}
    \Big(
    \aWW_{q\sigma,\chi^\prime\chi}^{(0)\text{sc}}
    +
    \aVV_{q\sigma,\chi^\prime\chi}^{(0)\text{sf}}
    \Big)
    -
    \Omega_{\chi^\prime\chi}
    \Big(
     \sum_{q\sigma}
    \eta_\sigma
    \eta_q
    \aVV_{q\sigma,\chi^\prime\chi}^{(0)\text{sf}}
    \Big)^{\!2}
    \Bigg]
    ,
    \\
    \label{Eq:L22}
    \kcL_{22}
    =\ &
    \frac{\Gamma}{T}
    \sum_{\chi\chi^\prime}
    \widetilde{\prob}_\chi
    \Bigg[
    \sum_{q\sigma}
    \Big(
    \aWW_{q\sigma,\chi^\prime\chi}^{(2)\text{sc}}
    +
    \aWW_{q\sigma,\chi^\prime\chi}^{(2)\text{sf}}
    +
    \frac{\Delta_{\chi\chi^\prime}^2}{4}
    \aVV_{q\sigma,\chi^\prime\chi}^{(0)\text{sf}}
    \Big)
    -
    \Omega_{\chi^\prime\chi}
    \Big(
    \sum_{q\sigma}
    \eta_q
    \Big\{
    \aWW_{q\sigma,\chi^\prime\chi}^{(1)\text{sf}}
    -
    \frac{\Delta_{\chi\chi^\prime}}{2}
    \aVV_{q\sigma,\chi^\prime\chi}^{(0)\text{sf}}
    \Big\}
    \Big)^2
    \Bigg]
    ,
    \\
    \label{Eq:L01_L10}
    \kcL_{01}
    =\ &
    \kcL_{10}
    =
    \frac{\Gamma}{T}
    \sum_{\chi\chi^\prime}
    \widetilde{\prob}_\chi
    \Bigg[
    \sum_{q\sigma}
    \eta_\sigma
    \aWW_{q\sigma,\chi^\prime\chi}^{(0)\text{sc}}
    -
    \Omega_{\chi^\prime\chi}
    \Big(
    \sum_{q\sigma}
    \eta_\sigma
    \eta_q
    \aVV_{q\sigma,\chi^\prime\chi}^{(0)\text{sf}}
    \Big)
    \!
    \Big(
    \sum_{q\sigma}
    \eta_q
    \aWW_{q\sigma,\chi^\prime\chi}^{(0)\text{sf}}
    \Big)
    \Bigg]
     ,
    \\
    \label{Eq:L02_L20}
    \kcL_{02}
    =\ &
    \kcL_{20}
    =
    \frac{\Gamma}{T}
    \sum_{\chi\chi^\prime}
    \widetilde{\prob}_\chi
    \Bigg[
    \sum_{q\sigma}
    \Big(
    \aWW_{q\sigma,\chi^\prime\chi}^{(1)\text{sc}}
    +
    \aWW_{q\sigma,\chi^\prime\chi}^{(1)\text{sf}}
    \Big)
    -
    \Omega_{\chi^\prime\chi}
    \Big(
    \sum_{q\sigma}
    \eta_q
    \aWW_{q\sigma,\chi^\prime\chi}^{(0)\text{sf}}
    \Big)\!
    \Big(
    \sum_{q\sigma}
    \eta_q
    \Big\{
    \aWW_{q\sigma,\chi^\prime\chi}^{(1)\text{sf}}
    -
    \frac{\Delta_{\chi\chi^\prime}}{2}
    \aVV_{q\sigma,\chi^\prime\chi}^{(0)\text{sf}}
    \Big\}
    \Big)
    \Bigg]
     ,
    \\
    \label{Eq:L12_L21}
    \kcL_{12}
    =\ &
    \kcL_{21}
    =
    \frac{\Gamma}{T}
    \sum_{\chi\chi^\prime}
    \widetilde{\prob}_\chi
    \Bigg[
    \sum_{q\sigma}
    \eta_\sigma
    \Big(
    \aWW_{q\sigma,\chi^\prime\chi}^{(1)\text{sc}}
    -
    \frac{\Delta_{\chi\chi^\prime}}{2}
    \aVV_{q\sigma,\chi^\prime\chi}^{(0)\text{sf}}
    \Big)
                        \nonumber\\
    &\hspace*{177pt}-
    \Omega_{\chi^\prime\chi}
    \Big(
     \sum_{q\sigma}
     \eta_\sigma
    \eta_q
    \aVV_{q\sigma,\chi^\prime\chi}^{(0)\text{sf}}
    \Big)\!
    \Big(
    \sum_{q\sigma}
    \eta_q
    \Big\{
    \aWW_{q\sigma,\chi^\prime\chi}^{(1)\text{sf}}
    -
    \frac{\Delta_{\chi\chi^\prime}}{2}
    \aVV_{q\sigma,\chi^\prime\chi}^{(0)\text{sf}}
    \Big\}
    \Big)
    \Bigg]
    ,
    \end{align}
\end{widetext}
and they satisfy the Onsager relation, $\kcL_{kn}=\kcL_{nk}$.~\cite{Onsager_Phys.Rev.37/1931_I,Onsager_Phys.Rev.38/1931_II,deGroot_book}
Note that the terms involving $\Omega_{\chi^\prime\chi}$, Eq.~(\ref{Eq:Omega}), describe corrections to the kinetic coefficients due to a deviation of the probability
distribution $\prob_\chi$  of the spin impurity states from the equilibrium distribution. Importantly, this deviation affects only contributions associated with spin-flip transitions.
The first interesting observation one can make about these coefficients concerns the fact of how they depend on two- ($\propto\aWW^{(n)\text{sc}}$ and $\propto\aWW^{(n)\text{sf}}$) and single-electrode ($\propto\aVV^{(0)\text{sf}}$) electron tunneling processes. Specifically, one can distinguish coefficients which even in the absence of the two-electrode processes (i.e., when $\aWW^{(n)\text{sc}}=\aWW^{(n)\text{sf}}=0$) can still remain nonzero, as for example $\kcL_{11}$, $\kcL_{22}$ and $\kcL_{12}$. Consequently, recalling the formulae for quantities describing thermoelectric properties of a system, discussed in Sec.~\ref{Sec:Overview_thermoelectrics}, one can expect the system under consideration to display a spin thermoelectric response without charge transport through the junction.

\section{\label{Sec:Results}Numerical results and discussion}

In the following, we focus on the analysis of  conventional and
spin-related  thermoelectric properties of the system of interest.
We distinguish three distinctive situations:
\begin{enumerate}
\item[(i)]
the tunnel junction in the absence of spin impurity
($\alpha_\text{ex}=0$), when only direct electron
tunneling between electrodes is possible ($\alpha_\text{d}=1$);
\item[(ii)]
the situation when the direct electron tunneling and the tunneling of
electrons with scattering on the impurity spin contribute comparably
(\mbox{$\alpha_\text{d}\approx\alpha_\text{ex}$}) to transport across
the barrier;
\item[(iii)]
the case with spin impurity ($\alpha_\text{ex}=1$), when
only single-electrode tunneling processes
($\alpha_\text{ex}^{LL}\neq0$ and $\alpha_\text{ex}^{RR}\neq0$) can occur, whereas
electron tunneling between two different electrodes is not admitted
($\alpha_\text{d}=\alpha_\text{ex}^{LR}=0$).
\end{enumerate}
In order to limit the parameter space to be investigated, for the general parameters describing tunneling of electrons and  the electronic band structure of the electrodes we assume:
$K=0.1$ eV, $\Lambda_L=\Lambda_R\equiv\Lambda=0.1$ eV$^{-3/2}$. In fact, since  $\kcL_{kn}\propto K^2\Lambda^2$ [cf. Eqs.~(\ref{Eq:notation_WV})  and~(\ref{Eq:L00})-(\ref{Eq:L12_L21}), and note that $\widetilde{\Phi}\propto\Lambda^2$], the values of $K$ and $\Lambda$ matter only for the magnitude of conductances~(\ref{Eq:G_def})-(\ref{Eq:kappa_def}), whereas thermopowers~(\ref{Eq:S_def}), and, consequently, also the Peltier coefficients~(\ref{Eq:Pi_def})  (not considered here) do not depend on these two parameters.
Furthermore, the spin impurity, if present, is usually coupled asymmetrically to the electrodes, and thus we assume $2\nu_L=\nu_R=1$, which means that in our case the impurity couples more strongly to the right electrode.
Finally, we assume that the spin number of the impurity is unaffected by the change of temperature in the considered range.

\subsection{Magnetic tunnel junction with no spin impurity}
Let us begin with the analysis of the
conceptually simplest case which corresponds to a bare magnetic tunnel
junction, i.e., the junction with no spin impurity, $\alpha_\text{ex}=0$. Assuming
$\alpha_\text{d}=1$, the relevant expressions for the kinetic
coefficients~(\ref{Eq:L00})-(\ref{Eq:L12_L21}) can be then written
in the following simple form:
     \begin{eqnarray}\label{Eq:L_alphaEX_0}
     \kcL_{nk}
     =
     \frac{2\Gamma}{T}
     \sum_{\sigma}
     \eta_\sigma^{n+k}
     \widetilde{\Phi}_{R\sigma}^{(\delta_{n2}+\delta_{k2})L\sigma}\!\big(0\big)
     ,
     \end{eqnarray}
where we made use of the fact that the function $\widetilde{\Phi}$ is symmetric under the
exchange of spin and electrode indices,
     $
\widetilde{\Phi}_{\overline{q}\sigma}^{(\delta_{n2}+\delta_{k2})q\sigma}\!\big(0\big)
     =
\widetilde{\Phi}_{q\sigma}^{(\delta_{n2}+\delta_{k2})\overline{q}\sigma}\!\big(0\big)
     $
[recall that $\widetilde{\Phi}\equiv\Phi\big|_\text{eq}$ and use
Eq.~(\ref{Eq:Phi_fun})].
One can immediately note that
     \begin{equation}
     \kcL_{nk}
     \propto
    \widetilde{\Phi}_{R\uparrow}^{(\delta_{n2}+\delta_{k2})L\uparrow}\!\big(0\big)
     -
    \widetilde{\Phi}_{R\downarrow}^{(\delta_{n2}+\delta_{k2})L\downarrow}\!\big(0\big)
    \end{equation}
if $n+k$ is an odd number.
This, in turn, leads to an important observation  that  the
spin-dependent transport quantities, such as the off-diagonal elements
$G^\text{m}$ and $G_\text{S}^\text{m}$ of the generalized
conductance matrix, Eq.~(\ref{Eq:G_def}), as well as the spin thermopower
$\mathcal{S}_\text{S}$, Eq.~(\ref{Eq:S_def}), depend on the
left-right electrode spin asymmetry \emph{via} the integrand factor
$\mathcal{A}(\omega)\equiv\rho_\uparrow^L(\omega)\rho_\uparrow^R(\omega)-\rho_\downarrow^L(\omega)\rho_\downarrow^R(\omega)$.
In particular, if the condition $\mathcal{A}(\omega)=0$ is fulfilled,
these quantities are identically equal to zero, and the junction does
not exhibit the spin thermoelectric response,
$\textrm{ZT}_\text{S}=0$.
This happens, for instance, when both electrodes are made of
the same ferromagnetic material (i.e., $E_L=E_R$ and $P_L=P_R$) and their
spin moments are oriented \emph{antiparallel}, so that
$\rho_\uparrow^L(\omega)=\rho_\downarrow^R(\omega)$ and
$\rho_\downarrow^L(\omega)=\rho_\uparrow^R(\omega)$.

\subsubsection{Generalized electrical conductance}

\begin{figure}[t]
    \includegraphics[width=0.95\columnwidth]{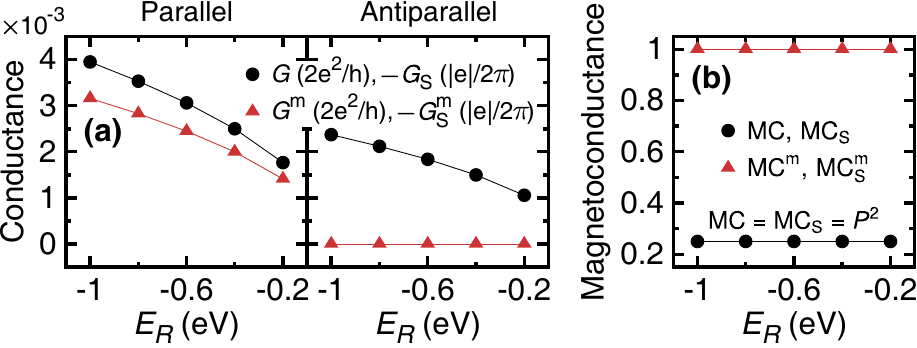}
     \caption{
     (color online)
     (a) Elements of the generalized conductance matrix $\vec{G}$ of a bare tunnel junction (i.e.,   in the absence of spin impurity, $\alpha_\text{ex}=0$) in the \emph{parallel}  (left) and \emph{antiparallel} (right) magnetic configuration, presented as a function of the position of the bottom edge of the conduction band in the right electrode, $E_R$, for a constant value of $E_L=-1$ eV, and for $P_L=P_R\equiv P=0.5$. The conductances are expressed in the units as indicated.
     The corresponding magnetoconductance is show in (b).
    Though the conductances are plotted for  $T=1$~K, they remain constant within the temperature range of interest, that is  up to $100$~K.
    }
     \label{Fig:Fig4}
\end{figure}

To illustrate basic thermoelectric characteristics of a bare junction,
especially their dependence on
the magnetic properties of electrodes (the spin polarization coefficient
$P_q$ and the band edge energy $E_q$) and on temperature, in
Fig.~\ref{Fig:Fig4}(a) we present the elements of the
conductance matrix $\vec{G}$ in the parallel and antiparallel magnetic configurations.
Due to the Onsager relations, $G_\text{S}^\text{m}=-(\hbar/2|e|)G^\text{m}$ (see Eq.~(\ref{Eq:G_def}) and the text below), so
$G_\text{S}^\text{m}$ and $G^\text{m}$
are equal in magnitude, but have opposite signs in the units used in Fig.~\ref{Fig:Fig4}(a). In the absence of spin impurity, also
the charge $G$ and spin $G_\text{S}$ conductances are equal in magnitude and have opposite signs
when expressed in the units as in Fig.~\ref{Fig:Fig4}(a). The latter equality
is not valid in the case with spin impurity, as will be seen later.
The results presented in Fig.~\ref{Fig:Fig4}(a) are for a symmetrical spin polarization of the electrodes,
$P_L=P_R\equiv P=0.5$.
However, the positions of the
band edges for the left and right electrodes, $E_L$ and $E_R$, are generally different.
Noteworthily, in order to keep a constant polarization factor, the Stoner splitting is modified accordingly, when the position of the band edge in one electrode is changed.

First of all, we note that the conductances shown in Fig.~\ref{Fig:Fig4}(a) stay
constant in the temperature range used below in Fig.~\ref{Fig:Fig5} for the thermal conductance and thermopowers.
Second, the nondiagonal conductances, $G_\text{S}^\text{m}$ and $G^\text{m}$, are vanishingly small
in the antiparallel configuration. Detailed analysis shows that
they vanish exactly only for
$E_L=E_R=-1$ eV, whereas for $E_R\neq E_L$ these conductances are nonzero,
albeit vanishingly small when compared with $G$
and $G_\text{S}$. The situation is different in the parallel configuration,
where $G_\text{S}^\text{m}$ and $|G^\text{m}|$ are comparable with
with $G$ and $|G_\text{S}|$. This behavior is a consequence  of the
symmetrical spin polarization. If $P_L\ne P_R$,
the conductances $G_\text{S}^\text{m}$ and $G^\text{m}$ are remarkable not only
in the parallel configuration, but also in the antiparallel one.

Difference in the conductance components in both magnetic configurations
can be used to define the corresponding magnetoconductance associated with the transition from parallel to antiparallel magnetic
configuration, as shown in Fig.~\ref{Fig:Fig4}(b).
The  magnetoconductances associated with $G$, $G_\text{S}$, $G_\text{S}^\text{m}$ and $G^\text{m}$
are  denoted in the following as ${\rm MC}$, ${\rm MC}_\text{S}$, ${\rm MC}_\text{S}^\text{m}$ and ${\rm MC}^\text{m}$, respectively, and are defined as
     \begin{equation}
     {\rm MX}
     =
     \frac{X_{\rm P}-X_{\rm AP}}{X_{\rm P}+X_{\rm AP}}
     ,
     \end{equation}
for $X=G$, $G_\text{S}$, $G_\text{S}^\text{m}$ and $G^\text{m}$,
where $X_{\rm P(AP)}$ denotes the corresponding conductance  in the parallel (antiparallel) magnetic configuration.
In general, this definition limits the magnetoconductance to the range $\langle-1,1\rangle$. In the case under consideration, the magnetoconductances shown in Fig.~\ref{Fig:Fig4}(b) are roughly constant. This follows from the fact that the leads' spin polarization is kept constant. Interestingly, as the magnetoconductances ${\rm MC}$ and ${\rm MC}_\text{S}$ are rather small and
${\rm MC}={\rm MC}_\text{S}=P^2$,
 the magnetoconductances associated with the nondiagonal elements, ${\rm MC}_\text{S}^\text{m}$ and ${\rm MC}^\text{m}$, are almost equal to the upper limit equal to~1. The latter is due to small values of the corresponding conductances in the antiparallel magnetic configuration, as already discussed above.

\subsubsection{Thermoelectric quantities: thermal conductance, thermopower and spin thermopower}

\begin{figure}[t]
    \includegraphics[width=0.8\columnwidth]{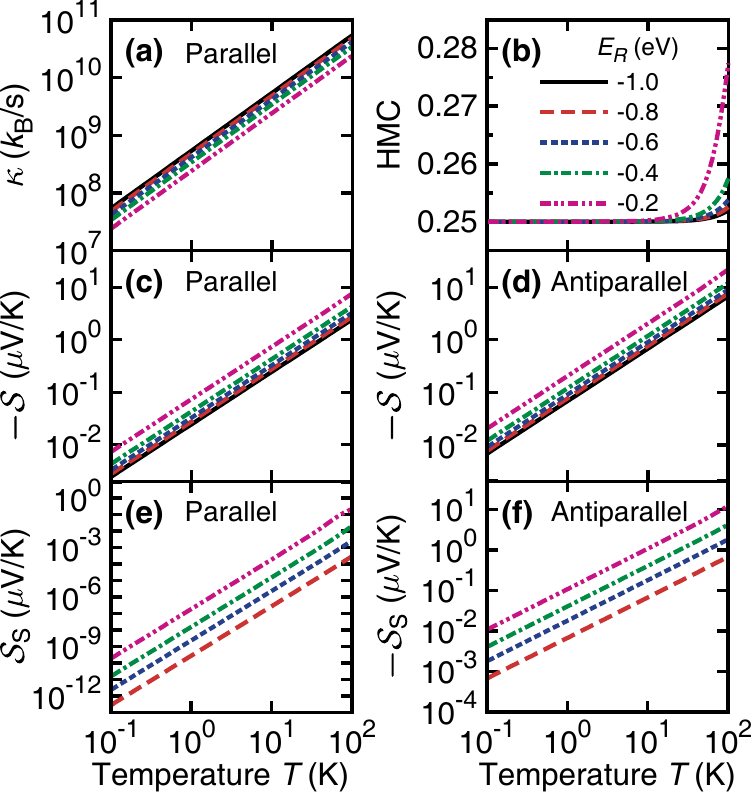}
     \caption{
     (color online)
     Thermal conductance (a), thermopower  $\mathcal{S}$ (c)-(d), and spin thermopower~$\mathcal{S}_\text{S}$ (e)-(f) calculated as a functions of temperature $T$ for indicated  values of $E_R$.
     Note that these values of $E_R$ correspond to the points indicted in Fig.~\ref{Fig:Fig4}.
     Whereas the heat conductance is presented for the parallel magnetic configuration, the thermopower and spin thermopower are  shown for both parallel and antiparallel configurations.
     The corresponding heat magnetoconductance  HMC is presented in (b).
    The other parameters as in Fig.~\ref{Fig:Fig4}.
    }
     \label{Fig:Fig5}
\end{figure}

Figure~\ref{Fig:Fig5} shows the temperature dependence of the
thermal conductance $\kappa$ (a), thermopower $\mathcal{S}$ (c)-(d), and spin thermopower $\mathcal{S}_\text{S}$ (e)-(f).  As one can see, the thermal conductance is a
linear function of temperature $T$. Closer numerical analysis proves
that the thermal conductance is related to the electrical one as
$\kappa=L_0GT$, with $L_0=\pi^2/(3e^2)$ denoting the Lorentz number
(recall that we set $k_\text{B}\equiv1$), which represents the
well-known Wiedemann-Franz law.~\cite{Barnard_book}
The relative change in the heat conductance, associated with the transition from parallel to antiparallel
magnetic configuration, is shown in  Fig.~\ref{Fig:Fig5}(b), where the  heat magnetoconductance (HMC)   have been defined similarly as the magnetoconductances considered above. The heat magnetoconductance increases only  slightly with temperature (note the logarithmic scale for the temperature).

A linear in temperature  behavior is also revealed  by the
thermopower  $\mathcal{S}$ and spin thermopower  $\mathcal{S}_\text{S}$, see Fig.~\ref{Fig:Fig5}(c)-(f),
where the solid line in (e)-(f) for
$E_R=-1$ is missing, as the spin thermopower is negligibly small in (e) and it vanishes exactly in (f).
First of all, one can observe that
the thermopower is negative for both magnetic configurations,
Fig.~\ref{Fig:Fig5}(c)-(d), which indicates that in both cases the corresponding thermocurrent is dominated
by particles. This follows from the particle-hole asymmetry in the DOS for the assumed model of electronic structure.
Conversely, the sign of the spin thermopower, Fig.~\ref{Fig:Fig5}(e)-(f), does depend on the magnetic configuration, which again is a consequence of  corresponding DOS in both spin channels.
The thermally-induced  spin current,
    \begin{equation}
    \hspace*{-3pt}
    I_\text{S}^\text{th}
    \equiv
    \frac{\hbar}{2T}\kcL_{12}\delta T
    \stackrel{\alpha_{\rm ex}=0}{=}
    \frac{\hbar}{2T}
    \Big[
    \widetilde{\Phi}_{R\uparrow}^{(1)L\uparrow}\!\big(0\big)
     -
    \widetilde{\Phi}_{R\downarrow}^{(1)L\downarrow}\!\big(0\big)
    \Big]
    \delta T
    ,
    \hspace*{-2pt}
\end{equation}
flowing in the parallel configuration is negative as the particle-hole asymmetry in the spin-minority (spin-down) channel is larger than in the spin-majority (spin-up) channel for the assumed parameters.
Accordingly, the spin voltage $\delta V_\text{S}$ required to compensate the spin current,
    \begin{equation}
    I_\text{S}\big|_{\delta V=0}
    =
    G_\text{S}\delta V_\text{S}+I_\text{S}^\text{th}
    \end{equation}
[with $G_\text{S}<0$ as shown in Fig.~\ref{Fig:Fig4}(a)],
is negative,
which explains why the corresponding spin thermopower is positive.
The situation is opposite in the antiparallel configuration, where the particle-hole asymmetry in the spin-up channel is larger. Thus, $I_\text{S}^\text{th}$ is positive and, in consequence,  the spin thermopower becomes negative.
Interestingly, one can notice that for the parallel
magnetic configuration, $\mathcal{S}_\text{S}$ takes extremely small values especially at
low temperatures, while $\mathcal{S}_\text{S}$ in the antiparallel configuration
is comparable in magnitude to $\mathcal{S}$ in both configurations.
However, we would like to emphasize that although the values of
thermopowers shown in Fig.~\ref{Fig:Fig5}(c)-(d) may
seem to be relatively small, they represent expected values for at
temperatures under consideration. For instance, in recent experiments on
the MgO-
\cite{Walter_NatureMater.10/2011,Liebing_Phys.Rev.Lett.107/2011} and
Al$_2$O$_3$-based \cite{LeBreton_Nature475/2011,Lin_NatureCommun.3/2012}
junctions it has been demonstrated that these systems can display at
room temperatures (RT) the thermopowers ranging from a few tens of $\mu$eV/K to
several mV/K, with typical values around~50-200~$\mu$eV/K.
On the other hand, for molecular break junctions based on
STM~\cite{Yee_NanoLett.11/2011,Evangeli_NanoLett.13/2013,Baheti_NanoLett.8/2008,
Malen_NanoLett.9/2009_1164,Widawsky_NanoLett.12/2011}  measured values of
thermopowers under the same conditions (RT) are usually smaller than  30~$\mu$eV/K, though theoretical predictions for some aromatic molecules~\cite{Bergfield_NanoLett.9/2009,
Bergfield_Phys.Rev.B79/2009}
are as large as 150~$\mu$eV/K.

Finally, an important conclusion of the present section is that large
values of thermopowers (and also figures  of merit, not shown)  are indeed associated with
large particle-hole asymmetry around the Fermi level, which here
follows from a significant difference in the band edge positions in
the two electrodes. For this reason, in the remaining part of the paper
we assume $E_L=-1$ eV and $E_R=-0.2$ eV.

\subsection{Transport in the presence of a spin impurity}

\begin{figure*}[t!!!]
    \includegraphics[width=1\textwidth]{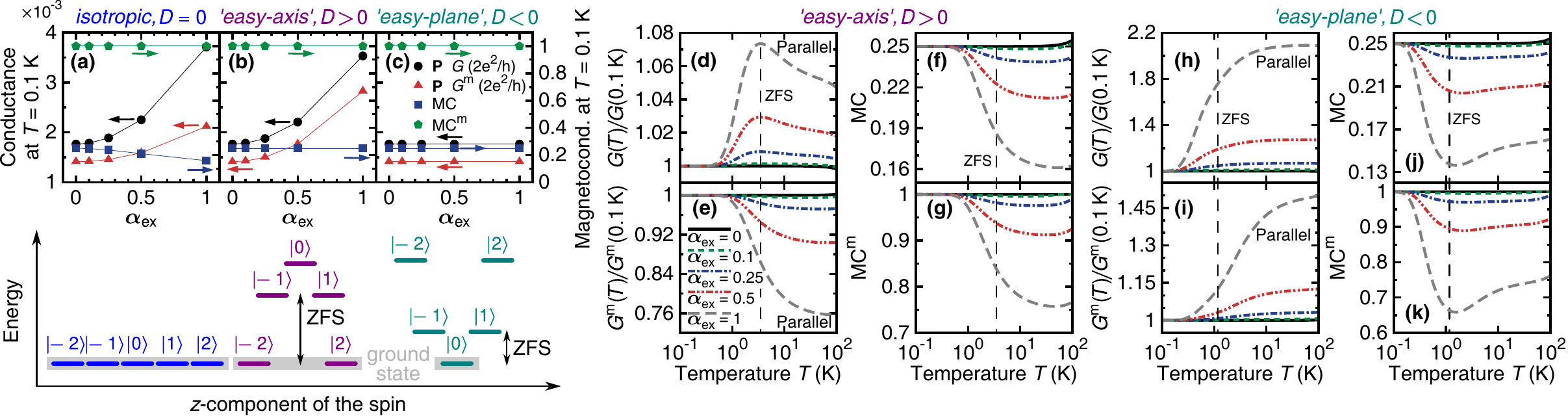}
     \caption{
     The effect of a spin impurity with $S=2$ on the electrical conductance of a magnetic tunnel junction.
     \emph{Left panel} [(a)-(c)]:
     Conductances $G$ and $G^\text{m}$ in the parallel (P) magnetic configuration, together with the corresponding magnetoconductances MC and MC$^\text{m}$, shown for $T=0.1$ K as a function of the parameter $\alpha_\text{ex}$ describing the electron tunneling with scattering on the impurity. Thin lines serve only as a guide for eyes.
     Three distinctive cases regarding the uniaxial magnetic anisotropy constant~$D$ (with $E=0$) are considered: (a) an isotropic impurity ($D=0$), and an anisotropic impurity with the spin anisotropy of (b) the `easy-axis' type ($D>0$), as well as (c) the `easy-plane' type ($D<0$) with $|D|=100$ $\mu$eV.
     Graphical depiction of the impurity energy spectrum for each of these three cases is shown below.
     \emph{Right panel} [(d)-(k)]:
     Dependence of the conductances in the parallel magnetic configuration, scaled to the corresponding values at $T=0.1$ K, and of the corresponding magnetoconductances on temperature for selected values of $\alpha_\text{ex}$ for $D>0$ (d)-(g) and $D<0$ (h)-(k).
     Vertical  dashed lines represent the temperature equivalent to the zero-field splitting: (d)-(g) $\text{ZFS}=3D\approx 3.5$ K and (h)-(k) $\text{ZFS}=|D|\approx 1.16$~K.
     We note that for $D=0$ the relevant conductances are constant within the temperature range under consideration.
     Other parameters: $\alpha_\text{d}=1$ and $P_L=P_R\equiv P=0.5$.
     }
     \label{Fig:Fig6}
\end{figure*}

Now, we analyze the effect of a spin impurity embedded in the barrier region of a magnetic tunnel junction on its thermoelectric properties, and start with examining how the electrical conductance of the junction changes with increasing the parameter~$\alpha_\text{ex}$ from $\alpha_\text{ex}=0$ (no impurity) to $\alpha_\text{ex}=1$, see Fig.~\ref{Fig:Fig6}.
As an example, we assume here a hypothetical impurity of an \emph{integer} spin $S=2$. The difference between the \emph{integer} and  \emph{half-integer} spin numbers will be explored later. We consider both  \emph{spin-isotropic} impurity ($D=E=0$) and \emph{spin-anisotropic} impurity, with the latter limited to the uniaxial magnetic anisotropy ($E=0$) of two different types: `easy-axis' ($D>0$) and `easy-plane' ($D<0$), cf. Fig.~\ref{Fig:Fig1}(b)-(c). For the purpose of this analysis, we assume the uniaxial anisotropy constant $|D|=100$ $\mu$eV, which is slightly larger that those observed for SMMs~\cite{Gatteschi_book,Mannini_Nature468/2010,Zyazin_NanoLett.10/2010} and some magnetic adatoms, e.g., Mn~\cite{Hirjibehedin_Science317/2007,Loth_NaturePhys.6/2010}, but smaller than for other adatoms, such as Co or Fe.~\cite{Hirjibehedin_Science317/2007,Brune_Surf.Sci.603/2009,Otte_NaturePhys.4/2008}

\subsubsection{Generalized electrical conductance}

To begin with, we note that the comparison of dimensionless magnitudes of the diagonal elements, $G/(2e^2/h)=h\kcL_{00}/2$ and $-G_\text{S}/(|e|/2\pi)=h\kcL_{11}/2$, of the generalized electrical conductance matrix $\vec{G}$, Eq.~(\ref{Eq:G_def}), shows that these quantities usually differ negligibly.  Therefore, for most of the present section only $G$ will be plotted.
Moreover, since the non-diagonal elements, $G^\text{m}$ and $G_\text{S}^\text{m}$,  are related as $G_\text{S}^\text{m}=-(\hbar/2|e|)G^\text{m}$, only~$G^\text{m}$ will be shown henceforth.
Because the contribution to the charge current due to the spin-conserving tunneling of electrons for $\alpha_\text{ex}\neq0$ depends on the impurity spin state, Eq.~(\ref{Eq:Wsc_def}), one expects the conductances to increase significantly as $\alpha_\text{ex}$ approaches~$\alpha_\text{d}$. Thus, in Fig.~\ref{Fig:Fig6} we first  present the conductances as a function of~$\alpha_\text{ex}$ (top panel) for one specific, low  temperature ($T=0.1$ K), and  then we examine the qualitative changes of the temperature evolution of conductances scaled to their low-temperature values for selected parameters $\alpha_\text{ex}$ (bottom panel).

One can see that the variation of the low-temperature conductances with increasing~ $\alpha_\text{ex}$, shown in Fig.~\ref{Fig:Fig6}(a)-(c) for the parallel magnetic configuration, depends strongly on the magnetic properties of the impurity spin. Whereas for $D=0$ (a) and $D>0$ (b) the conductances become larger with growing $\alpha_\text{ex}$, for $D<0$ (c) their values remain insensitive to $\alpha_\text{ex}$.
 Such a difference in the conductance behavior can be explained by considering the impurity spin states participating in scattering of conduction electrons traversing the barrier.
 In order to gain a deeper insight into the role of these states in electronic transport through a junction, we decompose the generalized conductance matrix as $\vec{G}=\vec{G}^\text{sc}+\vec{G}^\text{sf}$, where $\vec{G}^\text{sc}$ ($\vec{G}^\text{sf}$) represents the contribution corresponding to the tunneling processes during which the orientation of electronic spins is conserved (flipped).
The spin-conserving terms are then given by
    \begin{align}\label{Eq:G_sc}
    \hspace*{-3pt}
    \frac{G^\text{sc}}{2e^2/h}
    =
    -
    \frac{(G_\text{S})^\text{sc}}{|e|/2\pi}
    =
    \frac{h\Gamma}{T}
    \Big[
    &
    \alpha_\text{d}^2
    +
    \big(\alpha_\text{ex}^{LR}\big)^{\!2}
    \sum_{\chi}
    \widetilde{\prob}_\chi
    \big|\mathbb{S}_{\chi\chi}^z\big|^2
    \Big]
    \nonumber\\
    &
    \times
    \!\!
    \Big[
    \widetilde{\Phi}_{R\uparrow}^{(0)L\uparrow}\!\big(0\big)
    +
    \widetilde{\Phi}_{R\downarrow}^{(0)L\downarrow}\!\big(0\big)
    \Big]
    \!
    ,
    \hspace*{-2pt}
    \end{align}
and
    \begin{align}\label{Eq:Gm_sc}
    \hspace*{-3pt}
    \frac{(G^\text{m})^\text{sc}}{2e^2/h}
    =
    -
    \frac{(G_\text{S}^\text{m})^\text{sc}}{|e|/2\pi}
    =\ &
    \frac{h\Gamma}{T}
    \Big[
    \alpha_\text{d}^2
    +
    \big(\alpha_\text{ex}^{LR}\big)^{\!2}
    \sum_{\chi}
    \widetilde{\prob}_\chi
    \big|\mathbb{S}_{\chi\chi}^z\big|^2
    \Big]
    \nonumber\\
    &
    \hspace*{9pt}
    \times
    \!\!
    \Big[
    \widetilde{\Phi}_{R\uparrow}^{(0)L\uparrow}\!\big(0\big)
    -
    \widetilde{\Phi}_{R\downarrow}^{(0)L\downarrow}\!\big(0\big)
    \Big]
    \!
   .
   \hspace*{-2pt}
    \end{align}
Here, worthy of note is that as long as the transverse magnetic anisotropy is absent ($E=0$),
    $
    \sum_{\chi}
    \widetilde{\prob}_\chi
    \big|\mathbb{S}_{\chi\chi}^z\big|^2
    \equiv
    \sum_{\chi}
    \widetilde{\prob}_\chi
    \langle\chi|\hat{S}_z^2|\chi\rangle
    =
    \langle\hat{S}_z^2\rangle
    $.
One can also notice that the interference term,
    $
    \propto2
    \alpha_\text{d}\alpha_\text{ex}^{LR}
    \sum_{\chi}
    \widetilde{\prob}_\chi
    \mathbb{S}_{\chi\chi}^z
    \equiv
    2
    \alpha_\text{d}\alpha_\text{ex}^{LR}
    \langle\hat{S}_z\rangle
    $
[cf. Eq.~(\ref{Eq:Wsc_def})], is missing, since at equilibrium $\langle\hat{S}_z\rangle=0$.
On the other hand, the spin-flip terms take the form
    \begin{equation}\label{Eq:G_sf}
    \left\{
    \begin{aligned}
    &
    \frac{G^\text{sf(1)}}{2e^2/h}
    =
     \frac{h\Gamma}{2T}
    \sum_{\chi\chi^\prime}
    \widetilde{\prob}_\chi
    \sum_{q\sigma}
    \aWW_{q\sigma,\chi^\prime\chi}^{(0)\text{sf}}
    ,
    \\
    &
    \frac{G^\text{sf(2)}}{2e^2/h}
    =
    -
     \frac{h\Gamma}{2T}
    \sum_{\chi\chi^\prime}
    \widetilde{\prob}_\chi
    \Omega_{\chi^\prime\chi}
    \Big[
    \sum_{q\sigma}
    \eta_q
    \aWW_{q\sigma,\chi^\prime\chi}^{(0)\text{sf}}
    \Big]^{\!2}
    ,
    \end{aligned}
    \right.
    \end{equation}
    \begin{equation}\label{Eq:GS_sf}
    \left\{
    \begin{aligned}
    &
    \frac{(G_\text{S})^\text{sf(1)}}{|e|/2\pi}
    =
    -
    \frac{h\Gamma}{2T}
    \sum_{\chi\chi^\prime}
    \!
    \widetilde{\prob}_\chi
    \!
    \sum_{q\sigma}
    \!
    \aVV_{q\sigma,\chi^\prime\chi}^{(0)\text{sf}}
    ,
    \\
    &
    \frac{(G_\text{S})^\text{sf(2)}}{|e|/2\pi}
    =
    \frac{h\Gamma}{2T}
    \sum_{\chi\chi^\prime}
    \!
    \widetilde{\prob}_\chi
    \Omega_{\chi^\prime\chi}
    \Big[
     \sum_{q\sigma}
    \eta_\sigma
    \eta_q
    \aVV_{q\sigma,\chi^\prime\chi}^{(0)\text{sf}}
    \Big]^{\!2}
    ,
    \end{aligned}
    \right.
    \end{equation}
    \begin{align}\label{Eq:Gm_sf}
    \frac{(G^\text{m})^\text{sf}}{2e^2/h}
    =\ &
    -
    \frac{(G_\text{S}^\text{m})^\text{sf}}{|e|/2\pi}
    =
    -
    \frac{h\Gamma}{2T}
    \sum_{\chi\chi^\prime}
    \widetilde{\prob}_\chi
    \Omega_{\chi^\prime\chi}
    \nonumber\\
    &
    \hspace*{10pt}
    \times
    \Big[
    \sum_{q\sigma}
    \eta_\sigma
    \eta_q
    \aVV_{q\sigma,\chi^\prime\chi}^{(0)\text{sf}}
    \Big]
    \!
    \Big[
    \sum_{q\sigma}
    \eta_q
    \aWW_{q\sigma,\chi^\prime\chi}^{(0)\text{sf}}
    \Big]
    .
    \end{align}
Above, we have split the spin-flip contributions into two terms: the terms referred to as `$\text{sf(1)}$' present the main contributions due to spin-flip transitions, assuming the equilibrium probability of the spin impurity states, while the terms referred to as `$\text{sf(2)}$' involve $\Omega_{\chi^\prime\chi}$, Eq.~(\ref{Eq:Omega}), and describe corrections to the spin-flip contributions due to a deviation of the probability of spin states  from the equilibrium one.

\begin{figure}[t]
    \includegraphics[width=0.8\columnwidth]{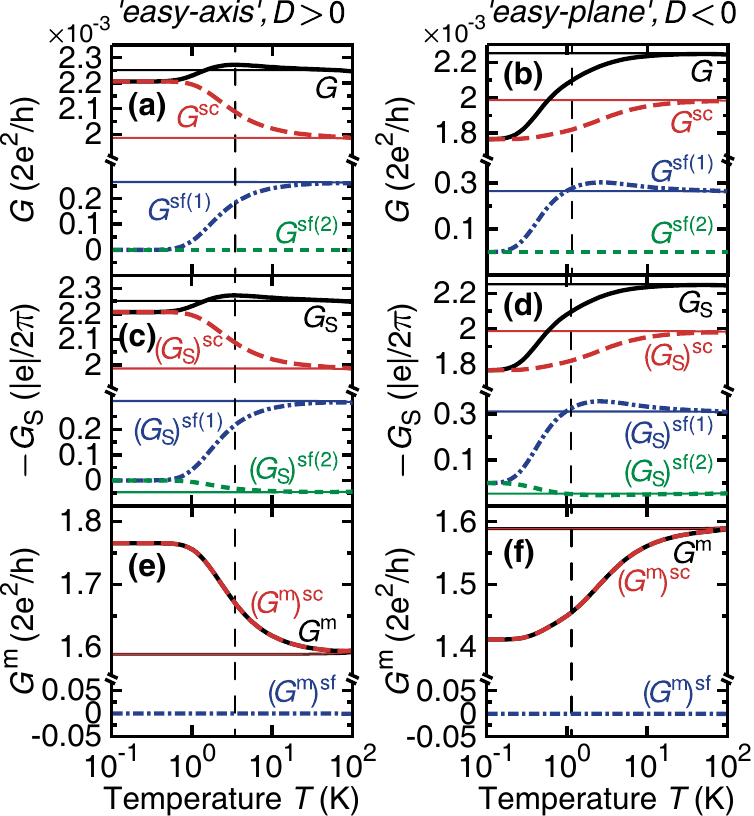}
     \caption{
     (color online)
     Decomposition of the diagonal, $G$ (a)-(b) and $G_\text{S}$ (c)-(d),  and non\-diagonal,~$G^\text{m}$ (e)-(f), elements of the generalized  conductance matrix $\vec{G}$ into spin-conserving (long-dashed lines marked as `sc') and spin-flip (short-dashed and dashed-dotted lines marked as `sf(1)' and
     `sf(2)') contributions shown as a function of temperature $T$ for $\alpha_\text{ex}=0.5$ and parallel magnetic configuration. Bold solid lines stand for the sum of all the contributions.
     Both the `easy-axis' ($D>0$, \emph{top panel}) and `easy-plane' ($D<0$, \emph{bottom panel}) types of the uniaxial magnetic anisotropy are analyzed.
     Furthermore, horizontal thin lines representing the spin-isotropic case ($D=0$) has been added for comparison.
     Note the break in the conductance scales, which has been introduced to highlight the qualitative changes of the contributions under consideration.
     Vertical lines, like in the right panel of Fig.~\ref{Fig:Fig6}, illustrate the relevant zero-field splittings.
     For detailed definitions of the contributions see Eqs.~(\ref{Eq:G_sc})-(\ref{Eq:Gm_sf}).
     All other parameters as in~Fig.~\ref{Fig:Fig6}.
     }
     \label{Fig:Fig7}
\end{figure}

In the spin-isotropic case, Fig.~\ref{Fig:Fig6}(a), all spin states are degenerate and can participate in the mechanism under discussion at any temperature, see the thin lines in Fig.~\ref{Fig:Fig7}.
In particular, from the equations above it is clear that the dominating contribution is the spin-conserving one due to the states $\ket{S_z=\pm S}$, that is the states of the largest $z$-component of the spin, though all the other states contribute as well.
 The situation changes in the presence of uniaxial magnetic anisotropy, which removes the degeneracy. For $D>0$, shown in ~Fig.~\ref{Fig:Fig6}(b), the ground state of the impurity is the doublet $\ket{S_z=\pm S}$, while all other intermediate states, $\ket{S_z}$ with $S_z=\pm S\mp1,\ldots,0$, have larger energies creating an energy barrier for spin reversal, and thus they do not take part in transport at low temperatures, i.e., for $T\ll\text{ZFS}$. In general,
ZFS stands for the \emph{zero-field splitting}, which basically represents the excitation energy between the ground and first excited states in the absence of an external magnetic field. Specifically, at present one gets $\text{ZFS}_{D>0}=(2S-1)D$, where the subscript `$D>0$' has been added to highlight the type of the uniaxial magnetic anisotropy the ZFS refers to. As a result, the lack of the contribution to transport from the intermediate impurity spin states manifests for $\alpha_\text{ex}\sim\alpha_\text{d}$ as slightly smaller values of the conductances when compared to the spin-isotropic case.
On the other hand, for $D<0$, shown in Fig.~\ref{Fig:Fig6}(c), the ground state of the impurity is the planar state $\ket{S_z=0}$, whose contribution  at low temperatures, $T\ll\text{ZFS}_{D<0}=|D|$, is identically equal to zero, see Eqs.~(\ref{Eq:G_sc})-(\ref{Eq:Gm_sc}). Thus,  the junction effectively behaves at low temperatures as in the absence of impurity, compare Fig.~\ref{Fig:Fig6}(c) with points for $E_R=-0.2$ eV in the left side of Fig.~\ref{Fig:Fig4}(a).

For a spin-anisotropic impurity, the effect of the excited spin states on transport can be observed at higher temperatures, $T\gtrsim\text{ZFS}$. This is illustrated in the bottom panel of Fig.~\ref{Fig:Fig6}, a well as in Fig.~\ref{Fig:Fig7} where different contributions to the conductances, Eqs.~(\ref{Eq:G_sc})-(\ref{Eq:Gm_sf}), are plotted separately for $\alpha_\text{ex}=0.5$.
Generally, one observes that as soon as temperature becomes of the order of the ZFS (marked by vertical dashed lines), the conductances  start deviating from their low-temperature values. Importantly, although the temperature evolution of these conductances depends qualitatively on the sign~of~$D$, the high-temperature (that is for $T\gg |D|S^2\approx4.6$ K, when all the impurity spin states contribute to transport) asymptotic values of $G$, $G_\text{S}$ and $G^\text{m}$, respectively, are the same for both types of the uniaxial magnetic anisotropy.
Moreover, these values also coincide with those for the spin-isotropic case.
For instance, compare the conductances at $T=100$ K in the top ($D>0$) and bottom ($D<0$) panels in Fig.~\ref{Fig:Fig7}.
Regarding the electrical conductance $G$, for $D>0$, Fig.~\ref{Fig:Fig6}(d), a maximum develops at $T\approx\text{ZFS}_{D>0}$, whereas for $D<0$, Fig.~\ref{Fig:Fig6}(h), only a monotonic increase of $G$ can be seen. The mechanism of the peak formation for $D>0$ originates from the spin-polarization of electron tunneling due to the presence of ferromagnetic electrodes, and the discussion of its details we defer to the end of this section. Nevertheless, one can still understand the dissimilar qualitative behavior of $G(T)$ for different signs of~$D$ by simply considering how the population of the impurity spin states changes with increasing $T$.

Let us consider the energy spectrum for $D>0$ in Fig.~\ref{Fig:Fig6}. Once $T$ becomes of the order of ZFS, the states $\ket{S_z=\pm1}$ become active in scattering of conduction electrons tunneling through the junction, and the conductance $G$ grows.
Interestingly,  this  growth is exclusively due to inelastic scattering of conduction electrons on the impurity, which are accompanied by the flip of electronic spins, see the component~$G^\text{sf(1)}$ in Fig.~\ref{Fig:Fig7}(a).  One can in general view this as the opening of new channels for transport.
On the other hand, the inclusion of excited impurity spin states with increasing $T$, and specifically the state of highest energy, $\ket{S_z=0}$, leads to a decrease  in the spin-conserving component~$G^\text{sc}$. Since at higher temperatures this decrease is not fully compensated by contribution due to  inelastic processes, one observes effectively a peak in the overall conductance $G$.
Moreover, the mechanism underlaying the reduction of~$G^\text{sc}$ becomes especially evident when considering the effect of the state~$\ket{S_z=0}$ on transport.  Although this state does not contribute directly to~$G^\text{sc}$, as the term~$\propto\big(\alpha_\text{ex}^{LR}\big)^{\!2}$  in Eq.~(\ref{Eq:G_sc}) vanishes, it still affects the conductance in the sense that its population probability builds up at the expense of the population probabilities of the other states. Thus, $G^\text{sc}$ decreases until all the states are populated with equal probabilities $1/(2S+1)$ at sufficiently high temperatures.
The opposite situation occurs for $D<0$, see the relevant energy spectrum in Fig.~\ref{Fig:Fig6}, where by increasing temperature one successively activates states $\ket{S_z=\pm1}$ and $\ket{S_z=\pm2}$. Importantly, the latter states when included are characterized by relatively modest population probabilities, but one should remember  that because of the large value of $S_z$ they contribute significantly to the magnitude of $G^\text{sc}$  through the term $\propto\big(\alpha_\text{ex}^{LR}\big)^{\!2}$  in Eq.~(\ref{Eq:G_sc}), see Fig.~\ref{Fig:Fig7}(b).
Analogous analysis can also be conducted for other elements of the generalized conductance matrix~$\vec{G}$.

At the beginning of this section we remarked that the calculated values of $G$ and~$G_\text{S}$ are approximately equal, which now can be explicitly seen if one compares the left ($G$) and middle ($G_\text{S}$) columns in Fig.~\ref{Fig:Fig7}. However, one should note that though the sum of all the contributions constituting either of the conductances are comparable, the values of specific  spin-flip contributions, e.g., $G^\text{sf(1)}/(2e^2/h)$ vs. $-(G_\text{S})^\text{sf(1)}/(|e|/2\pi)$, in each case are different. In particular, whereas $G^\text{sf(2)}$ is negligibly small, $(G_\text{S})^\text{sf(2)}$ is finite at large temperatures and has an opposite sign than $(G_\text{S})^\text{sf(1)}$.

To complete the discussion of charge/spin conducting properties of the junction with an impurity, we also analyze how they depend on the magnetic configuration. For this purpose, in Fig.~\ref{Fig:Fig6} apart from the conductances in the parallel configuration we also plotted the relevant magnetoconductances MC and~MC$^\text{m}$.
We observe that $\text{MC}=P^2$ (or $\text{MC}=0.25$ in the present case of $P=0.5$) for $T\ll\text{ZFS}$ and it gets reduced as soon as inelastic electron tunneling processes become thermally admitted. This tell us that although the junction electrically conducts better in the parallel configuration, $G_\text{P}>G_\text{AP}$, the difference is less pronounced at higher temperatures when the effect of the spin impurity is most significant.
A similar behavior occurs also for MC$^\text{m}$, but its low-temperature asymptotic value is 1, which means that $(G^\text{m})_\text{P}\gg(G^\text{m})_\text{AP}$ in the low-$T$ regime.

\subsubsection{Thermoelectric quantities: thermal conductance, thermopower and spin thermopower}

\begin{figure}[t]
    \includegraphics[width=0.95\columnwidth]{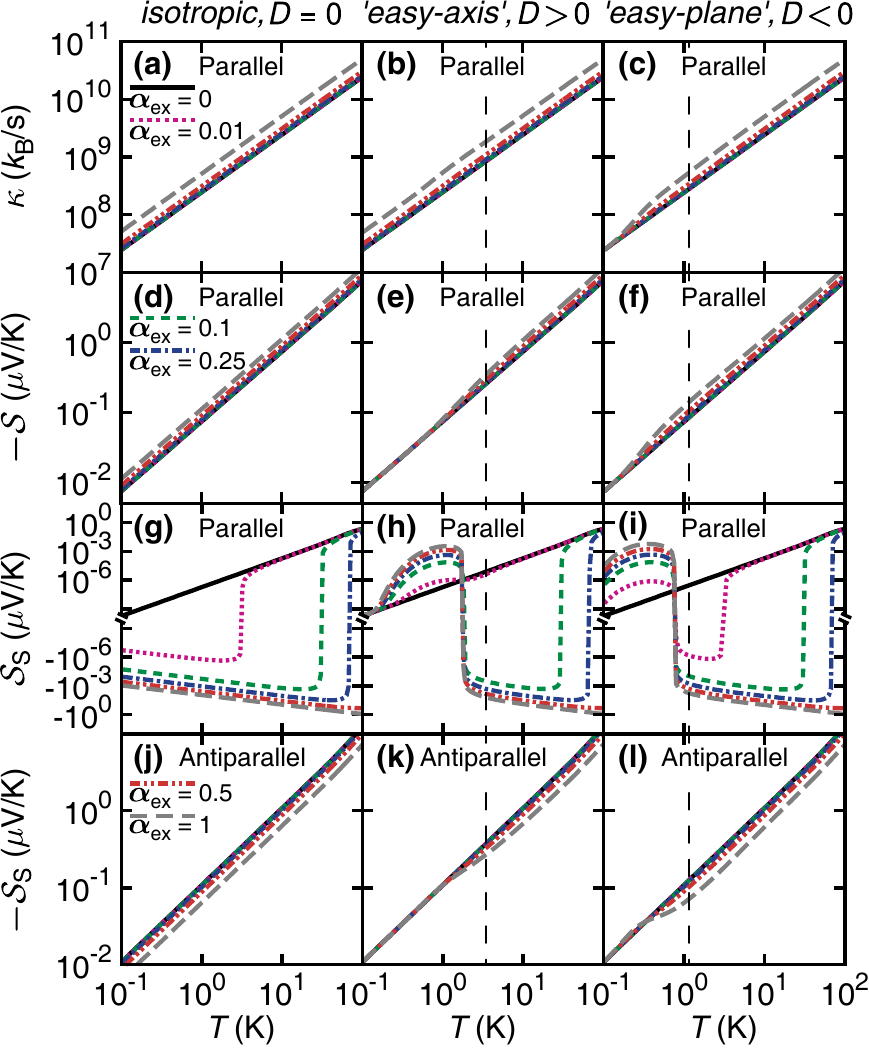}
     \caption{
     Thermal conductance $\kappa$ (a)-(c), thermopower~$\mathcal{S}$ (d)-(f) and spin thermo\-power~$\mathcal{S}_\text{S}$ (g)-(l) plotted as functions of temperature $T$ for several values of $\alpha_\text{ex}$.
     Similarly as in the left panel of Fig.~\ref{Fig:Fig6}, different columns correspond to (from left) $D=0$, $D>0$, and $D<0$.
     Unlike $\kappa$ and $\mathcal{S}$, which are monotonic functions of~$T$ in both  magnetic configurations, and thus only the parallel one is shown,
     $\mathcal{S}_\text{S}$ is a monotonic function of $T$ only in the antiparallel configuration (j)-(l), whereas in the parallel  one~(g)-(i) it is nonmonotonic and it can even change its sign with increasing $T$.
     Vertical lines, like in the right panel of Fig.~\ref{Fig:Fig6}, illustrate the relevant zero-field splittings.
     All parameters as in Fig.~\ref{Fig:Fig6}.
     }
     \label{Fig:Fig8}
\end{figure}

Before moving to the discussion of thermoelectric quantities, such as heat conductance $\kappa$, thermopower $\mathcal{S}$ and spin thermopower $\mathcal{S}_\text{S}$, shown in Fig.~\ref{Fig:Fig8}, it is worth recollecting that, unlike electrical/spin conductances, these depend nontrivially on the kinetic coefficients, cf. Eqs.~(\ref{Eq:kappa_def})-(\ref{Eq:S_def}). As a result, one cannot decompose them into the terms corresponding to the spin-conserving and spin-flip scattering processes, as it was done for the conductance matrix~$\vec{G}$.
In Fig.~\ref{Fig:Fig8}(a)-(c) we show how the temperature dependence of the heat conductance is modified when gradually turning on the interaction between tunneling electrons and the impurity, with the solid line representing the case of a bare junction ($\alpha_\text{ex}=0$) and the long-dashed line corresponding to the maximal effect of the impurity ($\alpha_\text{ex}=1$). Generally, as anticipated from the analysis of electrical conductance, the availability of the impurity spin states for scattering of electrons traversing the junction leads to increasing of energy transferred between the electrodes, which manifests as an increase in $\kappa$. Considering the kinetic coefficients $\kcL_{00}$, $\kcL_{22}$ and  $\kcL_{02}$, Eqs.~(\ref{Eq:L00}),(\ref{Eq:L22}) and (\ref{Eq:L02_L20}), which enter  the expression for $\kappa$, Eq.~(\ref{Eq:kappa_def}), it turns out that apart from two-electrode tunneling processes ($\propto\aWW^\text{(n)sc}$ and $\propto\aWW^\text{(n)sf}$ for $n=0,1,2$), also the inelastic single-electrode processes ($\propto\aVV^\text{(0)sf}$) should contribute. These processes are proportional to the transition energy $\Delta_{\chi^\prime\chi}$ between two states $\ket{\chi}$ and $\ket{\chi^\prime}$  due to scattering of a conduction electron on the impurity spin. Effectively, this is a new mechanism for energy transfer between the electrodes, which employs the impurity as an intermediate reservoir of energy. However, in the present situation, when electrons can freely traverse the junction, the contribution of energy transport by single-electrode tunneling processes seem to play a marginal role, as we explain below. For this reason, a more detailed discussion of such a mechanism has been postponed to Sec.~\ref{Sec:Single-electrode_tun_proc}, where spin thermoelectric effects  due to single-electrode tunneling will be explored.

Comparing the thermal conductance of the spin-isotropic impurity ($D=0$), Fig.~\ref{Fig:Fig8}(a), and the impurity with the `easy-axis' uniaxial spin anisotropy ($D>0$), Fig.~\ref{Fig:Fig8}(b), for $\alpha_\text{ex}\neq0$ one can hardly distinguish between these two cases. Moreover, for $D>0$ no change of $\kappa$ is seen at temperatures $T\approx\text{ZFS}$. This suggests that the observed, uniform increase of $\kappa$ within the analyzed temperature range occurs mainly due to the \emph{elastic} scattering of electrons on an impurity. On the other hand, for the impurity with the `easy-plane' spin anisotropy ($D<0$), Fig.~\ref{Fig:Fig8}(c), one can still notice the increase of $\kappa$ when $T$ approaches ZFS. This is a consequence of the fact that in such a case the ground state of the impurity is $\ket{S_z=0}$, which, similarly as for electrical/spin conductance,  does not contribute to elastic/spin-conserving transport.

Also for the conventional thermopower $\mathcal{S}$, Fig.~\ref{Fig:Fig8}(d)-(f), one observes an increase in $|\mathcal{S}|$ when $\alpha_\text{ex}$ approaches $\alpha_\text{d}$. Nevertheless, for $D>0$ a step appears for $T\approx\text{ZFS}$, which indicates that the growth in $|\mathcal{S}|$ can be attributed to spin-flip  electron scattering processes.
The negative sign of the thermopower indicates that the electron contribution is dominant in the whole temperature range in comparison to that due to holes.
The situation becomes more complex for the spin thermopower $\mathcal{S}_\text{S}$, Fig.~\ref{Fig:Fig8}(g)-(l). For the antiparallel magnetic configuration, Fig.~\ref{Fig:Fig8}(j)-(l), $|\mathcal{S}_\text{S}|$ displays a qualitative behavior opposite to that of~$|\mathcal{S}|$, i.e., $|\mathcal{S}_\text{S}|$ diminishes as $\alpha_\text{ex}$ is increased. The situation in the parallel configuration is different, Fig.~\ref{Fig:Fig8}(g)-(i), and the change of the thermopower sign can be observed.

To elucidate the origin of this behavior, let us consider the parallel configuration in more detail, starting from the isotropic  case, $D=0$.
For $\alpha_\text{ex}=0$, the spin thermopower is then positive in the entire temperature range studied in Fig.~\ref{Fig:Fig8}.
As described in the previous section,
this means that the
thermally-induced
spin current
$I_\text{S}^\text{th}$
flowing in the spin-minority channel is dominant,
which stems from the fact that
the particle-hole asymmetry in this channel is larger than that in the spin-majority channel.
Notably, for a finite
(relatively small) value of $\alpha_\text{ex}$, a transition at low temperatures from positive to negative spin thermopower takes place,
which is due to a reduction of $I_\text{S}^\text{th}$
in the spin-minority channel and increase in the spin-majority channel.
This appears  as a consequence of \emph{single-electrode spin-flip}  tunneling processes; see the second term (in brackets) of the spin current, Eq.~(\ref{Eq:I_S_def}), and terms $\propto\aVV^\text{(0)sf}$ in $\kcL_{12}$, Eq.~(\ref{Eq:L12_L21}). These processes are capable of modifying transport in spin channels by transferring an additional angular momentum through the junction \emph{indirectly}, i.e., \emph{via} the impurity -- see the next section for a detailed discussion.
The spin thermopower becomes positive again at larger temperatures. This transition, in turn, stems from  increased particle-hole asymmetry in the spin-minority channel as compared to that in the spin-majority one -- especially when the temperature becomes comparable to the Fermi energy in this spin channel (note we use $E_R=-0.2$ eV in the numerical calculations).
For $D\neq0$, the transition from positive to negative spin thermopower occurs when the thermal energy is comparable to ZFS, whereas the transition back to the positive spin thermopower takes place at temperatures comparable to those at which such a transition occurs for $D=0$.
On the other hand, in the antiparallel configuration (spin moment of the right electrode is reversed), $\mathcal{S}_\text{S}$ is negative and does not change sign in the temperature range of interest. Now the main contribution to $I_\text{S}^\text{th}$ comes from the spin-up
channel
and the
single-electrode spin-flip processes
for finite $\alpha_\text{ex}$ only enhance
the dominance of this channel,
so there is no sign change, and the absolute value  of the spin thermopower is larger than in the parallel magnetic state.

\subsubsection{The effect of transverse magnetic anisotropy}

\begin{figure*}[t!!!]
    \includegraphics[width=1\textwidth]{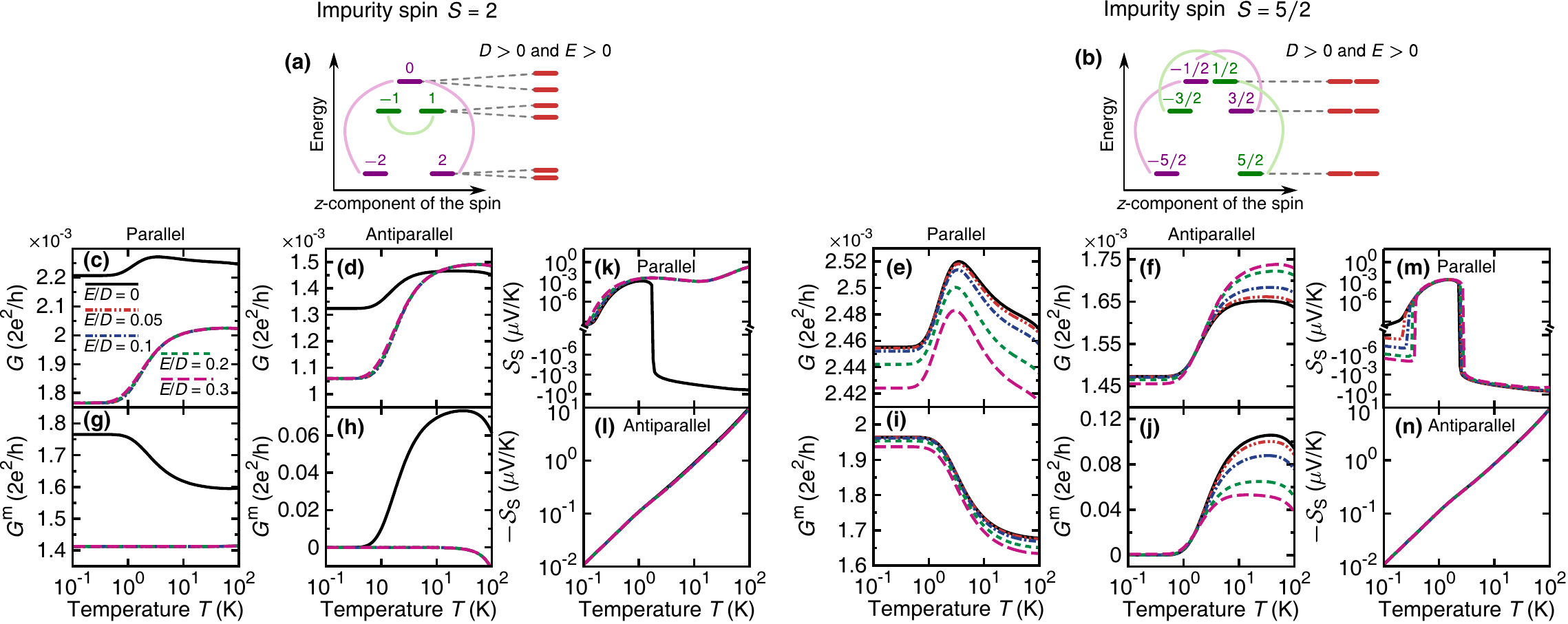}
     \caption{
     (color online)
     The influence of transverse magnetic anisotropy on the transport and thermoelectric properties of a magnetic tunnel junction with a spin impurity.
     Both  \emph{integer} ($S=2$, left side) and \emph{half-integer} ($S=5/2$, right side) impurity spin number are considered, and the uniaxial anisotropy of the `easy-axis' type is assumed \mbox{($D=100$~$\mu$eV)}.
     \emph{Top panel} [(a)-(b)]:
     Graphical illustration of the idea of how the transverse anisotropy term in the impurity Hamiltonian~(\ref{Eq:Ham_S}) mixes different states~$\ket{S_z}$ of the $z$-component of the impurity spin.
     \emph{Bottom panel} [(c)-(n)]:
     Elements of the generalized conductance matrix, $G$ (c)-(f) and $G^\text{m}$ (g)-(j), shown as a function of temperature $T$ for the parallel and antiparallel magnetic configurations.
     The solid lines illustrate the situation of~$E=0$.
     (k)-(n) Temperature dependence of the spin thermopower~$\mathcal{S}_\text{S}$  for the values of $E/D$ as indicated in (c).
     Note that both \emph{parallel} and \emph{antiparallel} magnetic configurations are considered.
     Other parameters: $\alpha_\text{d}=1$, $\alpha_\text{ex}=0.5$ and $P_L=P_R\equiv P=0.5$.
     }
     \label{Fig:Fig9}
\end{figure*}

So far we have focused on the situation when the magnetic properties of the impurity are dominated by its \emph{uniaxial} spin anisotropy. However, very often the uniaxial component of the anisotropy is also accompanied by the \emph{transverse} one, represented in Eq.~(\ref{Eq:Ham_S}) by the term proportional to $E$. Though, typical values of $E/|D|$ for SMMs~\cite{Mannini_Nature468/2010,Misiorny_DELFT} and magnetic adatoms\cite{Hirjibehedin_Science317/2007,Loth_NaturePhys.6/2010} are small, $E/|D|\lesssim0.2$, the presence of the  transverse spin anisotropy can have a profound effect on the system's properties. For instance, not only can it open the under-barrier quantum tunneling channels for spin reversal~\cite{Gatteschi_Angew.Chem.Int.Ed.42/2003} when $D>0$, but it also leads to clear manifestation of the geometric (or Berry-phase) effects in the spin dynamics.~\cite{Wernsdorfer_Science284/1999,Romeike_Phys.Rev.Lett.96/2006,
Romeike_Phys.Rev.Lett.96/2006_TranspSpetr,Leuenberger_Phys.Rev.Lett.97/2006,
Gonzalez_Phys.Rev.Lett.98/2007,Burzuri_Phys.Rev.Lett.111/2013}

In Fig.~\ref{Fig:Fig9} we show how the inclusion of the transverse spin anisotropy ($E\neq0$) affects the thermoelectric properties of the system under discussion. We plot the results for the case of $D>0$, and for the \emph{integer} ($S=2$, left panel) and \emph{half-integer} ($S=5/2$, right panel) impurity spin number. As already mentioned in Sec.~\ref{Sec:Setup}, the transverse anisotropy
causes mixing of the impurity spin states $\ket{S_z}$, so that $\hat{\mathcal{H}}_\text{imp}$ is no longer diagonal in the basis of such states. The new eigenstates~$\ket{\chi}$ are then combinations of the states $\ket{S_z}$. In particular, for an \emph{integer} $S$ each $\ket{\chi}$ is composed from the states coming from either of two uncoupled sets, $\big\{\forall\, S_z\in\mathbb{Z}_\text{even}:\ket{S_z}\big\}$ or
$\big\{\forall\, S_z\in\mathbb{Z}_\text{odd}:\ket{S_z}\big\}$,
while for a \emph{half-integer} $S$ from
$\big\{S_z=-S,-S+2,\ldots,-1/2,3/2,\ldots,S-3,S-1:\ket{S_z}\big\}$  or
$\big\{S_z=-S+1,-S+3,\ldots,-3/2,1/2,\ldots,S-2,S:\ket{S_z}\big\}$. Thus, according to the Kramers theorem,
all the degeneracy is thereby removed in the former case, while in the latter case it is preserved, since the two sets are time-reversed, cf. Fig.~\ref{Fig:Fig9}(a)-(b). Furthermore, we note that for an integer $S$ one gets $\langle\chi|\hat{S}_z|\chi\rangle=0$ for all states $\ket{\chi}$, whereas for a half-integer $S$ one still obtains  nonzero expectation values  of $\hat{S}_z$.

The above observation leads, in turn, to an important conclusion: for an impurity of integer spin the processes of spin-conserving scattering of conduction electrons on the impurity do not contribute to transport of charge, spin and energy.
In Fig.~\ref{Fig:Fig9} this is especially noticeable for the conductances $G$, (c)-(d), and $G^\text{m}$, (g)-(h); see also Eqs.~(\ref{Eq:G_sc}) and~(\ref{Eq:Gm_sc}) for $G^\text{sc}$ and $(G^\text{m})^\text{sc}$, respectively, where the term $\propto\big(\alpha_\text{ex}^{LR}\big)^{\!2}$ is identically equal to zero as soon as $E\neq0$. The constant value of $G^\text{m}$ in the temperature range of interest in (g)-(h) means that the charge (spin) current stimulated by a spin (electrical) bias arises exclusively due to direct tunneling of electrons through a junction, cf. Fig.~\ref{Fig:Fig7}(e) [$(G^\text{m})^\text{sf}$ is negligibly small also for $E\neq0$].
On the other hand, no significant qualitative changes of conductances are seen for a half-integer spin number ($S=5/2$), (e)-(f) and (i)-(j).
A similar behavior is observed for heat conductance $\kappa$ and thermopower $\mathcal{S}$ (not shown here), i.e., for an integer $S$ the transverse anisotropy only slightly diminishes the magnitude of $\kappa$, not affecting $\mathcal{S}$, whereas  for a half-integer $S$ both the quantities are hardly  influenced by a finite $E$.

The influence of transverse anisotropy on the spin thermopower $\mathcal{S}_\text{S}$ is more remarkable
in the parallel magnetic configuration, Fig.~\ref{Fig:Fig9} (k,m), whereas hardly any effect of $E\neq0$ can be seen in  the antiparallel configuration, Fig.~\ref{Fig:Fig9} (l,n).
This is due to modification of the spin eigenstates of the impurity by the transverse anisotropy. This modification, in turn, has a significant influence on the
single-electrode spin-flip processes mediating transfer of angular momentum across the junction  \emph{via} the impurity,
as well as on the spin-conserving part of the
two-electrode
contribution due to spin-impurity scattering.
For this reason, this modification  significantly impacts the thermally-induced spin current and, in consequence, also the spin thermopower.
Moreover, from the discussion above follows that
the effect of $E\neq0$
for half-integer $S$ is qualitatively different from that for integer $S$.
As can be seen in Fig.~\ref{Fig:Fig9}, the main influence of the transverse anisotropy on $\mathcal{S}_\text{S}$ in the parallel configuration  appears at low temperatures for half-integer $S$, where the transverse anisotropy leads to reversed sign of the spin thermopower,  and at higher temperatures for integer $S$, where the sign change due to single-electrode spin-flip transitions is suppressed.

\subsubsection{The effect of electrodes' spin polarization}

%
\begin{figure}[t]
    \includegraphics[width=0.8\columnwidth]{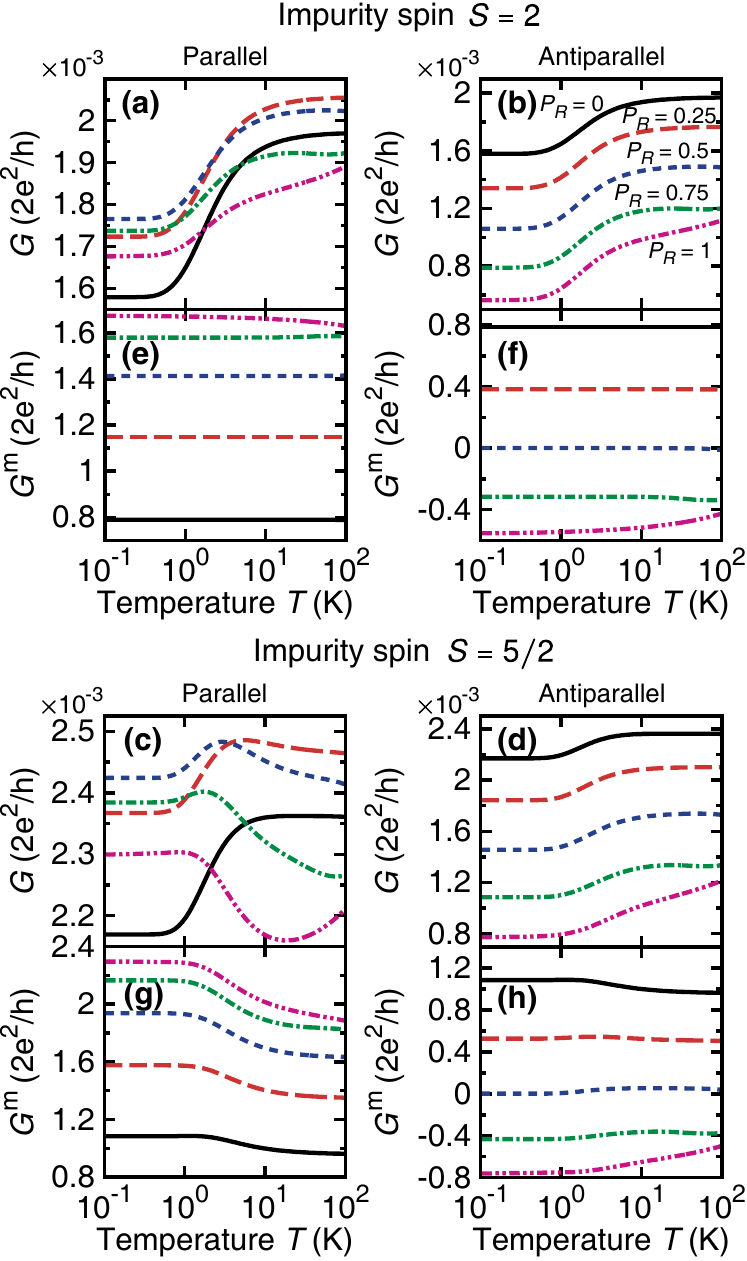}
     \caption{
     (color online)
     Electrical conductances $G$ (a)-(d) and $G^\text{m}$ (e)-(h) plotted vs. tem\-per\-a\-ture~$T$ for several values of the spin-polarization parameter $P_R$ of the right electrode and $P_L=0.5$.
     Similarly as in Fig.~\ref{Fig:Fig9}, two representative values of the impurity spin number are considered,  and the conductances are shown for the parallel and antiparallel magnetic configurations.
     Apart from $E/D=0.3$, all other parameters are assumed the same as in Fig.~\ref{Fig:Fig9}.
     }
     \label{Fig:Fig10}
\end{figure}
\begin{figure}[t]
    \includegraphics[width=0.8\columnwidth]{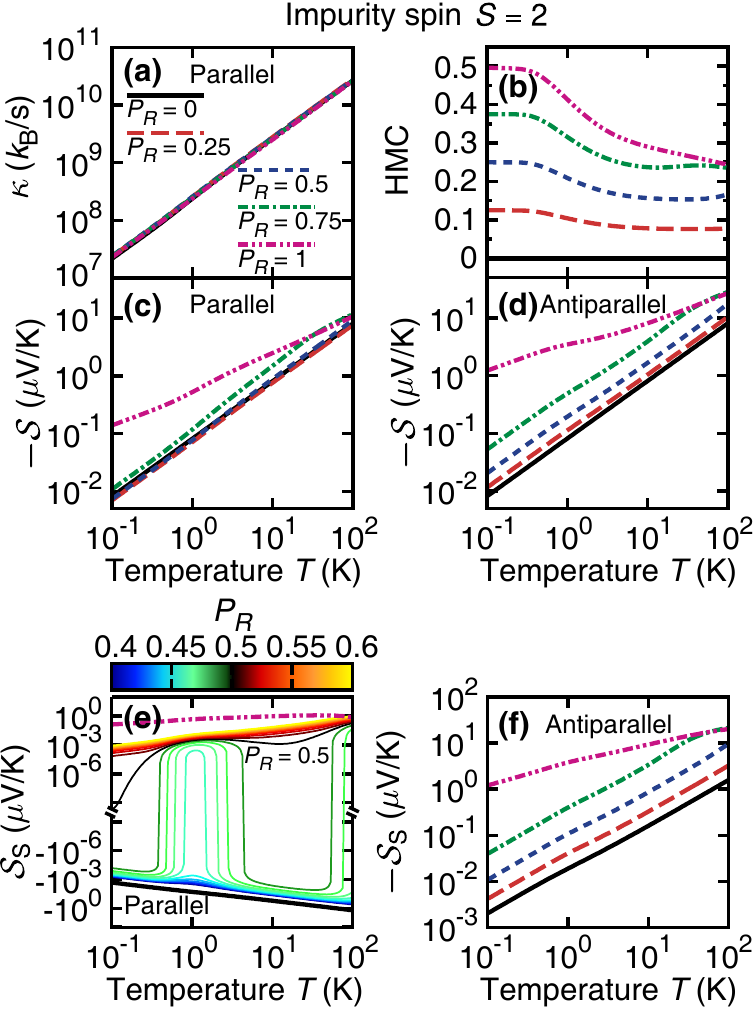}
     \caption{
     (color online)
     Analogous to Fig.~\ref{Fig:Fig10}, except that now the heat conductance~$\kappa$~ (a) and the corresponding heat magnetoconductance HMC (b), together with  the thermopower~$\mathcal{S}$~(c)-(d)  and spin thermopower~$\mathcal{S}_\text{S}$ (e)-(f) are shown.
     Whereas $\kappa$ is plotted only for the parallel magnetic configuration, $\mathcal{S}$ and $\mathcal{S}_\text{S}$ are presented for both parallel and antiparallel configurations.
     \emph{Bottom panel:}
     In order to trace the sign change of the spin thermopower in (e) from the negative one for small $P_R$ (see bold solid line for the limit of $P_R=0$) to the positive one  for large $P_R$ (see dashed-double-dotted line for the limit of $P_R=1$), in the range $P_R=\big\langle0.4,0.6\big\rangle$ with the interval $\Delta P_R=0.01$ we plot a series of curves (thin lines). The value of $P_R$ for each of these curves is color-coded, with a black line representing $P_R=0.5$.
     }
     \label{Fig:Fig11}
\end{figure}

Finally, before we conclude this section,  let us analyze how the spin polarization of electrodes influences the spin-dependent thermoelectric properties of a magnetic tunnel junction with a spin impurity.
Up to this point, we have been discussing a very specific situation,  when both the electrodes are characterized by the same spin-polarization parameter, $P_L=P_R\equiv P$.
Let us now relax this constriction by assuming only the spin polarization of the left electrode to be fixed, $P_L=0.5$, whereas the spin polarization of the right electrode $P_R$ can be varied between~0 (a nonmagnetic electrode) to~1
(a~half-metallic electrode).
The corresponding results are shown in Figs.~\ref{Fig:Fig10} and~\ref{Fig:Fig11}.

Considering first the conductances $G$ and $G^\text{m}$, Fig.~\ref{Fig:Fig10}, the effect of the difference in spin polarizations of the electrodes occurs mainly by modification of the spin-conserving contributions $G^\text{sc}$ and $(G^\text{m})^\text{sc}$, see Eqs.~(\ref{Eq:G_sc}) and~(\ref{Eq:Gm_sc}), respectively,  \emph{via} the $\widetilde{\Phi}$-functions, Eq.~(\ref{Eq:Phi_fun}). As one can notice, in the presence of transverse magnetic anisotropy, the temperature dependence of $G$ and $G^\text{m}$ for $S=2$ does not change qualitatively upon varying $P_R$. In the parallel magnetic configuration, the maximal electrical conductance $G$ in  (a) is observed only when $P_L$ and $P_R$ differ slightly. On the contrary, in the antiparallel configuration, the maximal value of $G$ in (b) occurs as $P_R\rightarrow0$, and $G$  decreases monotonically with increasing $P_R$ in the entire temperature range under consideration.
Moreover, the conductance $G^\text{m}$ for the parallel configuration~(e)  is the largest (and positive) for fully polarized right electrode, $P_R=1$. In the antiparallel configuration shown in (f),  $G^\text{m}$ vanishes for a symmetrical situation, $P_R=P_L$, and becomes negative when $P_R>P_L$ (the largest negative $G^\text{m}$ appears for $P_R=1$).
Note that for $P_R=0$ the distinction between the parallel and antiparallel magnetic configurations becomes irrelevant as the right electrode is then nonmagnetic, so that the solid lines in Fig.~\ref{Fig:Fig10}(a) and (b) [and analogously the ones in Fig.~\ref{Fig:Fig10}(e) and (f)] are actually equivalent.

On the other hand, for $S=5/2$ all the conductances, except $G$ in the parallel magnetic configuration, Fig.~\ref{Fig:Fig10}(c), exhibit analogous general response to the change of $P_R$ as for $S=2$.
Analyzing the shape of the electrical conductance curves plotted in Fig.~\ref{Fig:Fig10}(c) we can now address the question of the possible origin of the peak observed in Fig.~\ref{Fig:Fig7}(a). Indeed, it can be noticed that its occurrence is related to a specific magnetic configuration of electrodes, that is a clear peak is formed only if $P_L$ and $P_R$ are comparable and the spin moments of the electrodes are parallel.
Although the dependence  of $G^\text{m}$ on  the spin-polarization $P_R$ is qualitatively similar to that for $S=2$, and hence it is not discussed here,  one feature deserves a comment, i.e.,  a much more pronounced variation of $G^\text{m}$ with $T$ in comparison to a very weak temperature dependence of $G^\text{m}$ for $S=2$. This is related to a different role of perpendicular anisotropy in the integer $S$ and half-integer $S$ situations, as already discussed above, see Fig.~\ref{Fig:Fig9}.

Finally, variation of the heat conductance, thermopower and spin thermopower with polarization $P_R$ of the right electrode is shown in
Fig.~\ref{Fig:Fig11}. Since the dependence on $P_R$ for $S=5/2$ is qualitatively similar to that in the case of $S=2$, in this figure we present results only
for the latter case. The heat conductance $\kappa$ in (a) is roughly independent of $P_R$, and increases
almost linearly
with temperature. The corresponding heat magnetoconductance, HMC in (b), varies weakly with temperature and decreases with decreasing $P_R$. Obviously, the HMC vanishes exactly in the limit of nonmagnetic right electrode, $P_R=0$.
Qualitatively similar dependence on $P_R$ can be also observed for the thermopower $\mathcal{S}$, cf. (c) and (d).  In turn, the spin thermopower $\mathcal{S}_\text{S}$, shown in~(e)-(f), depends on $P_R$ in a more complex manner, which follows from the fact that in the parallel magnetic configuration it can change sign with increasing temperature, as already discussed before, see Figs.~\ref{Fig:Fig8} and~\ref{Fig:Fig9}.

\subsection{\label{Sec:Single-electrode_tun_proc}Spin-dependent thermoelectric effects in the absence of charge transport across the junction}

It has been shown in the previous section that if tunneling of electrons through the junction is allowed, the corresponding two-electrode processes ($\propto\aWW^\text{(n)sc}$ and $\propto\aWW^\text{(n)sf}$) determine the transport characteristics of the system. In particular, the main contribution to the charge, spin and energy transport stems from spin-conserving elastic tunneling processes, when conduction electrons either tunnel directly between the two electrodes or they are scattered elastically  by the impurity when traversing the junction, but neither energy nor angular momentum is exchanged between electrons and the impurity, see, e.g., Fig.~\ref{Fig:Fig7} for charge $G$ and spin $G_\text{S}$ conductances.
Though the single-electrode tunneling processes ($\propto\aVV^\text{(0)sf}$) do not play the dominant role in the presence of two-electrode processes, they become of crucial importance in the case when the latter ones are suppressed,~\cite{Ren_Phys.Rev.B89/2014,Misiorny_Phys.Rev.B89/2014}
and their prominent role for transport of spin and energy will be now discussed.

\begin{figure}[t!!!]
    \includegraphics[width=0.8\columnwidth]{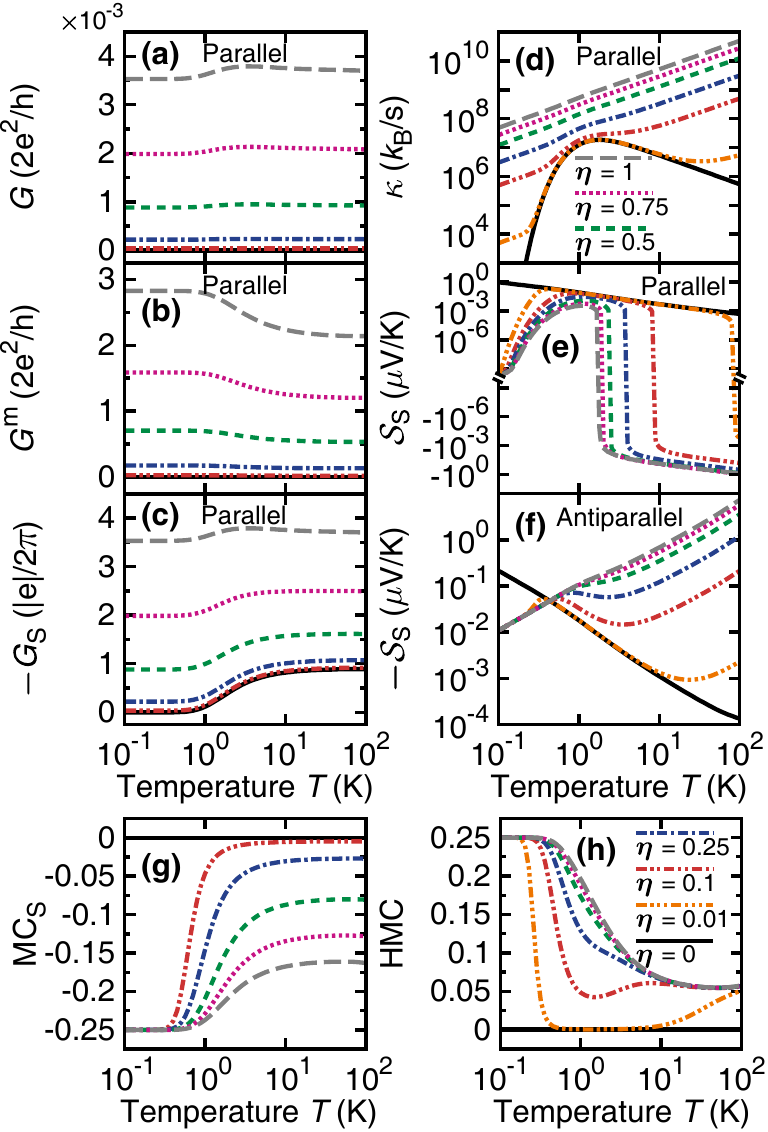}
     \caption{
     (color online)
     The effect of switching off the two-electrode tunneling processes on ther\-mo\-electric characteristics of a model system with an impurity of spin $S=2$ (with $D=100$~$\mu$eV and $E=0$).
     \emph{Top panel} [(a)-(f)]:
     The \emph{left column} presents the temperature dependence~of~$G$~(a),  $G^\text{m}$ (b), and spin $G_\text{S}$ (c) conductances in the parallel magnetic configuration for several values of the parameter $\eta$ (see the main text). %
     In the \emph{right column}, heat conductance~$\kappa$ (d) in the parallel magnetic configuration  and spin thermopower $\mathcal{S}_\text{S}$ (e)-(f) in both the configurations are plotted.
     The \emph{bottom panel} shows the spin magnetoconductance MC (g) and heat magnetoconductance HMC~(h).
     Other parameters: $\alpha_\text{d}=\alpha_\text{ex}=1$ and $P_L=P_R=0.5$.
     }
     \label{Fig:Fig12}
\end{figure}

\begin{figure}[t!!!]
    \includegraphics[width=0.9\columnwidth]{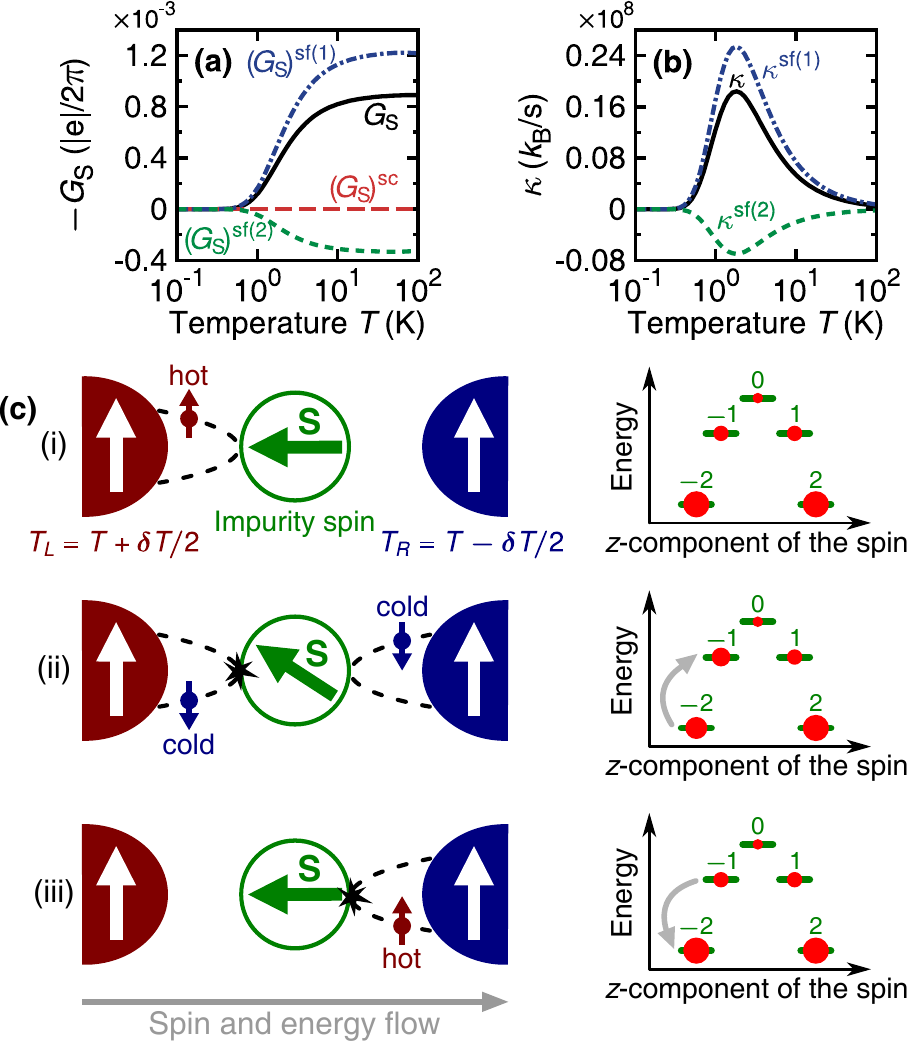}
     \caption{
     (color online)
     Decomposition of spin $G_\text{S}$ (a) and heat  $\kappa$ (b) conductances into different contributions, see Eqs.~(\ref{Eq:GS_sf}) and (\ref{Eq:kappa_sf}), respectively, shown as functions of temperature $T$ for $\eta=0$, with all other parameters as in Fig.~\ref{Fig:Fig12}.
     In (c) an example mechanism based on single-electrode electron tunneling precesses, that leads to transport of spin and energy through the junction
    is  depicted.
    Recall that the left electrode has higher temperature than the right one, $T_L>T_R$.
    The rightmost panel represents energy spectra of the spin impurity with the occupation probabilities for all spin states schematically marked with the dots.
     }
     \label{Fig:Fig13}
\end{figure}

To begin with let us  assume again (for the sake of simplicity) a junction with the spin impurity of $S=2$ exhibiting only the uniaxial magnetic anisotropy ($E=0$), and analyze how the thermoelectric properties of such a  system change when it becomes electrically
insulating,
i.e., the transfer of electrons between the left and right electrodes is blocked. For this purpose, we use the following substitution: $\alpha_\text{d}\rightarrow\eta\alpha_\text{d}$ and $\alpha_\text{ex}^{LR}\rightarrow\eta\alpha_\text{ex}^{LR}$.
The dimensionless parameter $\eta$ quantifies the presence of the two-electrode tunneling processes, with $\eta=0$ ($\eta=1$) representing the lack (maximum effect) of such processes.
Figure~\ref{Fig:Fig12} illustrates the evolution of transport characteristics of a junction with a spin impurity for $\eta$ decreasing from $\eta=1$ (long-dashed lines) to $\eta=0$ (solid lines). It can be seen that whereas the electrical conductances $G$~(a) and $G^\text{m}$~(b) approach 0 as $\eta\rightarrow0$
regardless of temperature,
the spin conductance $G_\text{S}$~(c) still
attains finite
values at high temperatures, $T\gtrsim\text{ZFS}$, and $G_\text{S}=0$ otherwise.
This means that even though charge is not transferred across the junction, one can still observe the flow of spin.
In order to understand better this phenomenon in Fig.~\ref{Fig:Fig13}(a)  we plot separately the contributions~$(G_\text{S})^\text{sc}$, $(G_\text{S})^\text{sf(1)}$ and $(G_\text{S})^\text{sf(2)}$ for $\eta=0$, cf. Eqs.~(\ref{Eq:G_sc}) and~(\ref{Eq:GS_sf}).

First of all, we note that when  the two-electrode tunneling processes are suppressed ($\eta=0$), only
single-electrode
scattering events leading to reversal of the electronic spin contribute to the spin current, Eq.~(\ref{Eq:GS_sf}).
In Fig.~\ref{Fig:Fig13}(c) we schematically present how the processes under consideration can result in spin transport between electrodes without charge being transferred.
For example, a conduction electron coming from the electrode of higher temperature (i) scatters on the impurity spin, which leads to the flip of the electron's spin orientation~(ii). Since angular momentum must be conserved in the system, it means that the quantum of angular momentum $\hbar$ has been exchanged with the impurity. In particular, for a spin-up incoming electron considered in (i), the electron has delivered it to the impurity. Then, a similar process can occur between the other electrode (of lower temperature) and the impurity (ii)-(iii). If as a result of this scattering process, an electron now changes its spin orientation so that it subtracts angular momentum from the impurity (iii), effectively a quantum of angular momentum will be transported through the junction.
Importantly, because the exchange of angular momentum between electrons and the impurity requires excitation of the latter, thus~$G_\text{S}$ differs significantly from zero only if $T$ becomes of the order of  ZFS, see Fig.~\ref{Fig:Fig13}(a). Moreover, unlike in the case when also the two-electrode tunneling process are active, Fig.~\ref{Fig:Fig7}(c), where the contribution of $(G_\text{S})^\text{sf(2)}$ to $G_\text{S}$ is marginal, at present the effect of processes contributing to $(G_\text{S})^\text{sf(2)}$ is important.

Interestingly, the above analysis leads to a conclusion that apart from angular momentum, an electron scattering on the impurity should also in general exchange energy with it. This comes as a direct consequence of the presence of uniaxial magnetic anisotropy ($D\neq0$).
For this reason, transport of energy in the situation under discussion should take place as well, which can be seen in Fig.~\ref{Fig:Fig12}(d) as a non-zero thermal conductance $\kappa$ for $\eta=0$ (solid line). Similarly as in the case of  spin conductance,
Eq.~(\ref{Eq:GS_sf}), also the thermal conductance can be
at present decomposed as $\kappa=\kappa^\text{sf(1)}+\kappa^\text{sf(2)}$, with
    \begin{equation}\label{Eq:kappa_sf}
    \hspace*{-3pt}
    \left\{
    \begin{aligned}
    &
    \kappa^\text{sf(1)}
    =
    \frac{\Gamma}{T^2}
    \sum_{\chi\chi^\prime}
    \widetilde{\prob}_\chi
    \frac{\Delta_{\chi\chi^\prime}^2}{4}
    \sum_{q\sigma}
    \aVV_{q\sigma,\chi^\prime\chi}^{(0)\text{sf}}
    ,
    \\
    &
    \kappa^\text{sf(2)}
    =
    -
    \frac{\Gamma}{T^2}
    \sum_{\chi\chi^\prime}
    \widetilde{\prob}_\chi
    \frac{\Delta_{\chi\chi^\prime}^2}{4}
    \Omega_{\chi^\prime\chi}
    \Big[
    \sum_{q\sigma}
    \eta_q
    \aVV_{q\sigma,\chi^\prime\chi}^{(0)\text{sf}}
    \Big]^2
    ,
    \end{aligned}
    \right.
    \end{equation}
where now [cf. Eq.~(\ref{Eq:Omega})]
    \begin{equation}
    \Omega_{\chi^\prime\chi}
    =
    \Big\{
    \sum_{q\sigma}
    \aVV_{q\sigma,\chi^\prime\chi}^{(0)\text{sf}}
    \Big\}^{-1}
    ,
    \end{equation}
see Fig.~\ref{Fig:Fig13}(b).
Analogously as in the case of $(G_\text{S})^{\text{sf(2)}}$, the contribution $\kappa^\text{sf(2)}$ to the heat conductance represents the effect of a deviation of the probability distribution of the impurity spin states from the equilibrium one. One can see that the inclusion of $\kappa^\text{sf(2)}$ is essential for a proper description of energy transport in the situation under discussion.

Another point worthy of note is that $G_\text{S}$ and $\kappa$ do not depend on the magnetic configuration for $\eta =0$. As a result, the corresponding spin magnetoconductance MC$_\text{S}$ and heat magnetoconductance $\text{HMC}$ vanish exactly. This is rather clear as the probability of process contributing to $G_\text{S}$ and $\kappa$ on the left side depend on the product of DOS for majority and minority electrons, similarly as on the right side. Since such products are independent on magnetic configuration,  MC$_\text{S}=0$ and  $\text{HMC}=0$.

Since both transport of spin and energy can  in principle arise  in the system even if no transport of charge is permitted, one can thus expect the system to exhibit a finite spin thermopower as well.
Taking into account the general expression
 for thermo-kinetic coefficients,  in the limit of $\eta=0$ one finds the following formula for
$\mathcal{S}_\text{S}$:
\begin{widetext}
    \begin{equation}
    \mathcal{S}_\text{S}
    =
    \frac{1}{|e|T}
    \!
    \cdot
    \!
    \frac{
    \sum_{\chi\chi^\prime}
    \widetilde{\prob}_\chi
    \frac{\Delta_{\chi\chi^\prime}}{2}
    \Big\{\!
    \sum_{q\sigma}
    \eta_\sigma
    \aVV_{q\sigma,\chi^\prime\chi}^{(0)\text{sf}}
    -
    \Omega_{\chi^\prime\chi}
    \Big[
     \sum_{q\sigma}
     \eta_\sigma
    \eta_q
    \aVV_{q\sigma,\chi^\prime\chi}^{(0)\text{sf}}
    \Big]\!
    \Big[
    \sum_{q\sigma}
    \eta_q
    \aVV_{q\sigma,\chi^\prime\chi}^{(0)\text{sf}}
    \Big]
    \!
    \Big\}
    }{
    \sum_{\chi\chi^\prime}
    \widetilde{\prob}_\chi
    \Big\{\!
    \sum_{q\sigma}
    \aVV_{q\sigma,\chi^\prime\chi}^{(0)\text{sf}}
    -
    \Omega_{\chi^\prime\chi}
    \Big[
     \sum_{q\sigma}
    \eta_\sigma
    \eta_q
    \aVV_{q\sigma,\chi^\prime\chi}^{(0)\text{sf}}
    \Big]^{\!2}
    \Big\}
    }
    .
    \end{equation}
\end{widetext}
It can be seen in Fig.~\ref{Fig:Fig12}(e)-(f)  that the spin thermopower is positive in the parallel magnetic configuration, while in the antiparallel configuration it is negative. This can be accounted for by considering  detailed balance of spin reversal process on both left and right sides of the junction. In the parallel configuration the associated
thermally-induced
spin current $I_\text{S}^\text{th}$ is negative and the induced spin voltage is also negative, so the spin thermopower is positive. When the
magnetic moment
of the right electrode is reversed,
$I_\text{S}^\text{th}$  is also reversed  and, consequently, the spin thermopower changes sign becoming negative.

\subsubsection{The effect of uniaxial and transverse\\ magnetic anisotropy}

\begin{figure}[t!!!]
    \includegraphics[width=0.8\columnwidth]{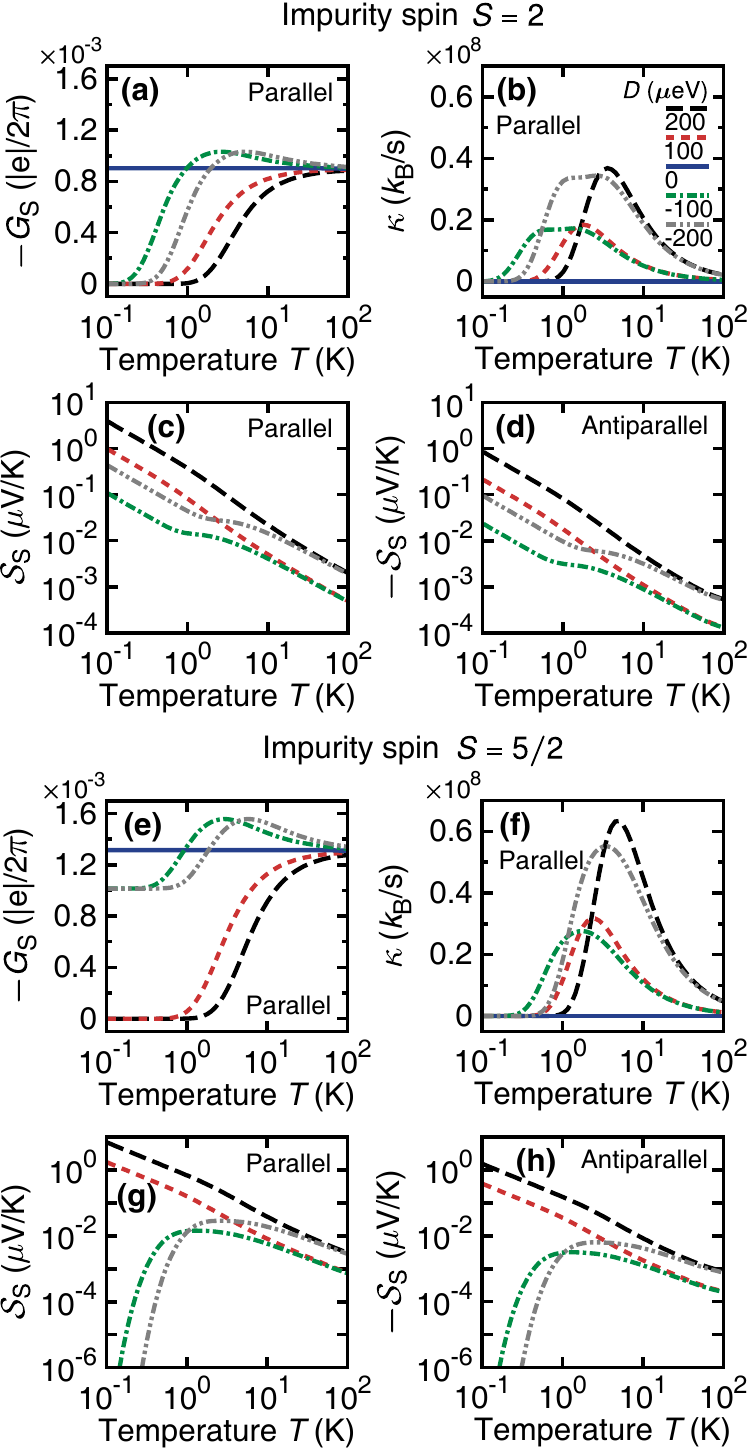}
     \caption{
     (color online)
     Analysis of the influence of the uniaxial magnetic anisotropy (quantified by~$D$) on the spin-dependent thermoelectric characteristics of the system in the absence of the charge transfer between electrodes.
     Two different cases of the impurity spin number $S$ are presented:  for
      $S$ being \emph{integer} (a)-(d) and
      for $S$ being \emph{half-integer} (e)-(h).
     Spin~$G_\text{S}$ (a,e) and heat $\kappa$ (b,f) conductances in the parallel magnetic configuration, as well as spin thermopower $\mathcal{S}_\text{S}$ both in the parallel (c,g) and antiparallel (d,h) configurations are plotted as a function of temperature $T$.
     Note that the solid line represents here the case of a \emph{spin-isotropic} impurity ($D=0$), whereas  the dashed (dashed-dotted) lines refer to a \emph{spin-anisotropic} impurity with the uniaxial anisotropy of the `easy-axis' (`easy-plane') type.
     Moreover, in the bottom panel $(\mathcal{S}_\text{S})_\text{P}=(\mathcal{S}_\text{S})_\text{AP}=0$ for $D=0$.
     Remaining parameters: $\eta=0$, $\alpha_\text{ex}=1$, $E=0$ and $P_L=P_R=0.5$.
     }
     \label{Fig:Fig14}
\end{figure}

As we discussed above, the uniaxial magnetic anisotropy seems to be  crucial for the occurrence of spin and energy transport, as well as spin thermopower. To further investigate this point, in Fig.~\ref{Fig:Fig14} we plot the thermoelectric quantities in question for both \emph{integer} ($S=2$) and \emph{half-integer} ($S=5/2$) values of the spin number of the impurity, and we consider  the uniaxial magnetic anisotropy of the `easy-axis' ($D>0$, dashed lines)  and `easy-plane' ($D<0$, dashed-dotted lines) type as well as the spin-isotropic case ($D=0$, solid line).
First of all, we observe that even though  transport of spin is possible for $D=0$ ($G_\text{S}\neq0$),  energy cannot be transferred due to the degeneracy of the impurity spin states, which reveals as $\kappa=0$ and $\mathcal{S}_\text{S}=0$ at any temperature.
Second, both the value of the uniaxial magnetic anisotropy constant $|D|$ and the spin number $S$  determine the maximal attainable value of the thermal conductance. It stems from the fact that these two parameters define the excitation energy between two neighboring spin states, e.g., for the transition between the states $\ket{S_z=\pm S}$ and $\ket{S_z=\pm S\mp 1}$ one has the excitation energy of $(2S-1)|D|$, cf. Fig.~\ref{Fig:Fig13}(b) and~(f).
Third, for a half-integer $S$ and $D<0$ the ground state of the impurity is the doublet $\ket{S_z=\pm1/2}$, so that even at low temperature $G_\text{S}$ has a non-zero value.

\begin{figure*}[t]
    \includegraphics[width=0.65\textwidth]{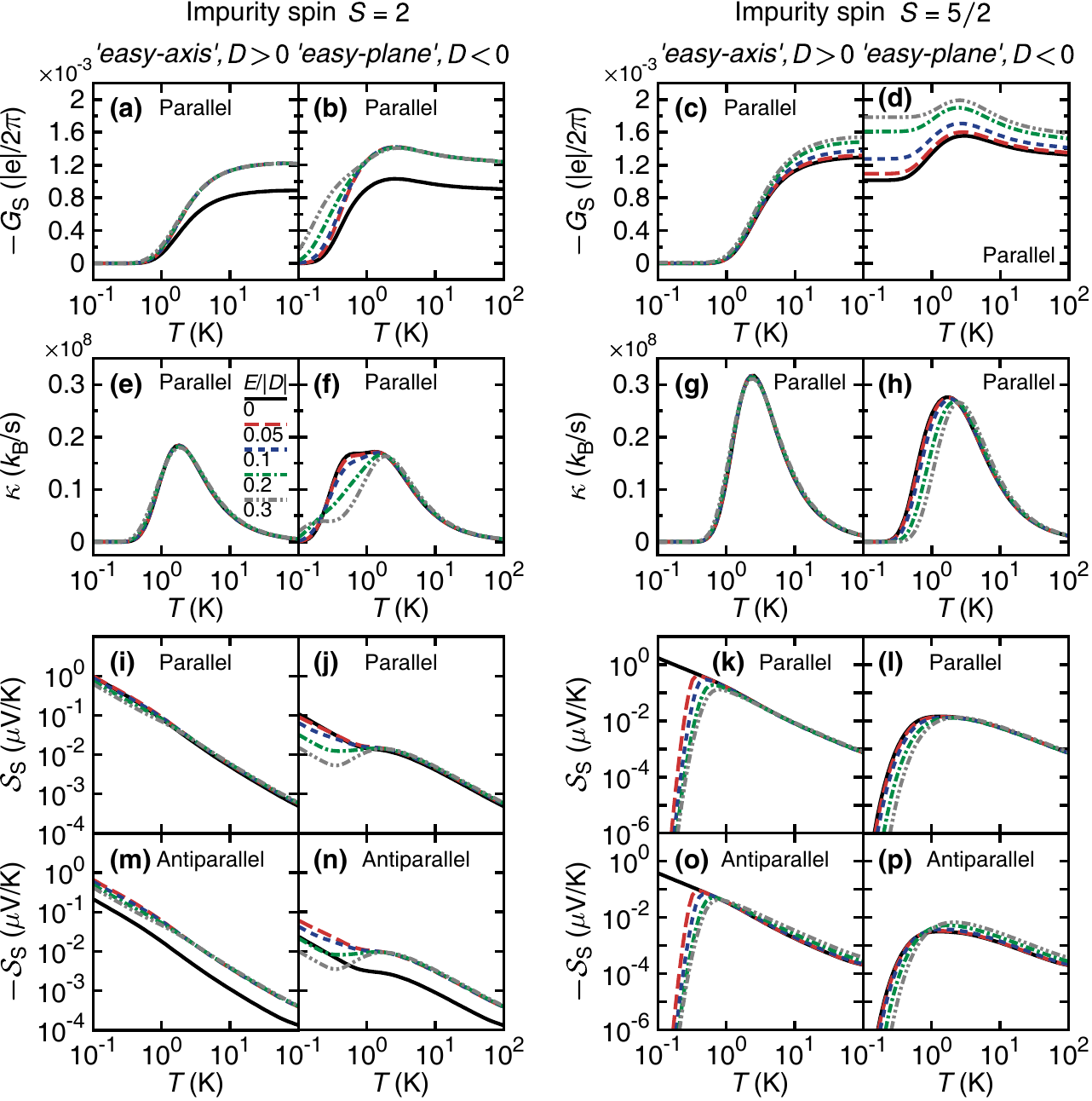}
     \caption{
     (color online)
     Analogous to Fig.~\ref{Fig:Fig14}, but now the effect of transverse magnetic anisotropy $E$ is investigated for two different types of the uniaxial magnetic anisotropy, $D>0$ and $D<0$. Except $|D|=100$ $\mu$eV, all other parameters as in Fig.~\ref{Fig:Fig14}.
     }
     \label{Fig:Fig15}
\end{figure*}

A significant qualitative difference in the behavior of spin conductance and spin thermopower for integer and half-integer $S$ appears for negative $D$. This difference is due to the fact that for half-integer spin and $D<0$ there is no barrier for spin-flip transition at low temperatures, while for $D>0$ (similarly as for integer $S$) such a barrier does exist. Accordingly, the spin conductance for half-integer $S$ is finite at low temperatures for $D<0$ and it vanishes  for $D>0$. In turn, the spin thermopower  is significantly reduced at low temperatures for $D<0$ and $S=5/2$ in comparison to the spin thermopower for integer $S$. Interestingly, the spin thermopowers in the parallel and antiparallel configurations behave in a qualitatively similar way with temperature. In both cases the spin thermopower is positive in the parallel configuration and negative in the antiparallel one. As before, this is a consequence of the  fact that in the  parallel configuration the dominant contribution to the spin current is from spin-down electrons, while in the antiparallel one  the dominant spin current flows in the spin-up channel.

Behavior of all thermoelectric and transport parameters changes when the effects of transverse magnetic anisotropy are included, Fig.~\ref{Fig:Fig15}.
These modifications, however, are remarkable mainly at low temperatures, while at higher temperatures they are much less pronounced.
Physical origin of these modifications
lies in the  change
the spin-impurity states
undergo owing to the presence of
the transverse anisotropy, as already discussed above.
However, one point is worth emphasizing, namely the pronounced effect of $E\neq0$ on the spin thermopower of an impurity with half-integer $S$, see Fig.~\ref{Fig:Fig15}(k,o). The significant decrease of $\mathcal{S}_\text{S}$ by several orders of magnitude, which  occurs at low temperatures, results from removing the energy barrier for the impurity spin reversal by the transverse anisotropy, as for $E\neq0$ direct transitions between the ground state doublet $\ket{S_z=\pm S}$ are allowed. In such a case, the system behaves qualitatively in a similar way for both $D>0$ and $D<0$, compare plots (k) and (o) with (l) and (p), respectively.

\subsubsection{The effect of electrodes' spin polarization}

\begin{figure}[t!!!]
    \includegraphics[width=0.8\columnwidth]{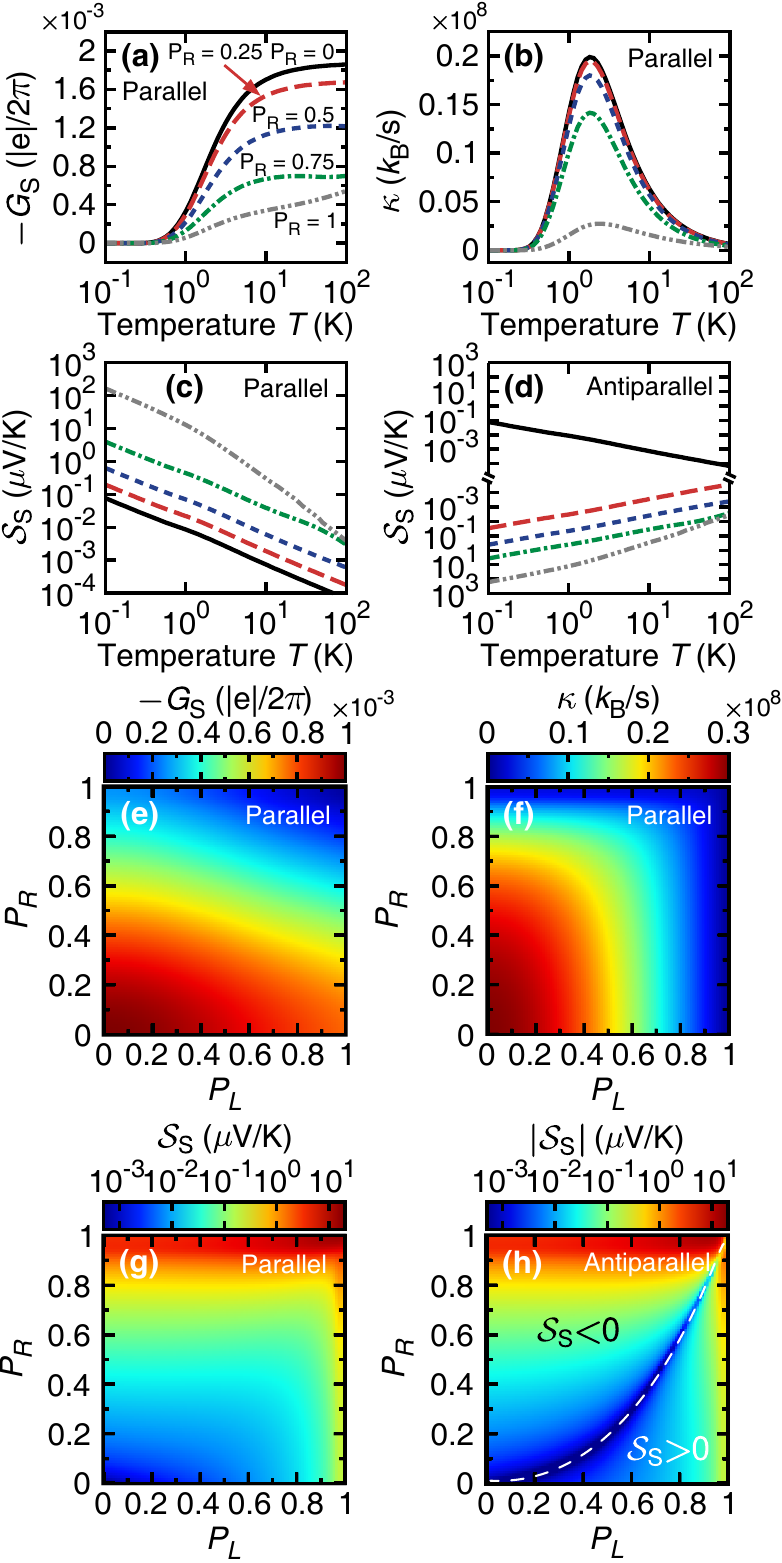}
     \caption{
     (color online)
     Temperature evolution of the spin conductance $G_\text{S}$ (a), heat conductance~$\kappa$~(b), and spin thermopower $\mathcal{S}_\text{S}$ (c)-(d) shown in the case of  the junction including the spin impurity with $S=2$ ($D=100$ $\mu$eV and $E/D=0.3$) for selected values of the spin-polarization parameter of the right electrode $P_R$ ($P_L=0.5$).
     All the quantities in question are presented for the parallel magnetic configuration, and $\mathcal{S}_\text{S}$ additionally also in the antiparallel one.
     \emph{Bottom panel} [(e)-(h)]:
     Analogous to the top panel, except that now the quantities under discussion are plotted as functions of  the spin polarizations of both electrodes, $P_L$ and $P_R$, for $T=2$ K.
     Dashed line in (h) separates visually the region of negative $\mathcal{S}_\text{S}$ (above the line) from the region of positive $\mathcal{S}_\text{S}$ (below the line).
     Other parameters: $\eta=0$ and $\alpha_\text{ex}=1$.
     }
     \label{Fig:Fig16}
\end{figure}

To complete the discussion of transport of spin and energy exclusively due to  single-electrode electron tunneling processes, we finally comment how this is affected by asymmetry in the electrodes' spin polarizations $P_L$ and $P_R$. Such a dependence is studied in Fig.~\ref{Fig:Fig16}, where it can be seen that a proper choice of $P_L$ and $P_R$ is of key importance for enhancing the effects under discussion. In particular, by setting $P_R=0$, which corresponds to a nonmagnetic right electrode, one can significantly increase the magnitudes of spin $G_\text{S}$ (a) and heat $\kappa$ (b) conductances. Since these two quantities depend now only on the product of spin-up and spin-down DOS for each of  electrodes separately, check up on  terms $\aVV_{q\sigma,\chi^\prime\chi}^{(0)\text{sf}}$ in Eqs.~(\ref{Eq:GS_sf}) and~(\ref{Eq:kappa_sf}) [also recall Eqs.~(\ref{Eq:notation_WV}) together with~(\ref{Eq:Phi_fun})], one actually expects  that the largest values of $G_\text{S}$ and $\kappa$ are to be obtained when  $P_L$ and $P_R$ are small but not necessarily equal, see Fig.~\ref{Fig:Fig16}(e)-(f).
This is not the case for spin thermopower $\mathcal{S}_\text{S}$, for which an opposite trend is observed, Fig.~\ref{Fig:Fig16}(c)-(d) and~(g)-(h), namely $\mathcal{S}_\text{S}$ reaches its maximal value when at least one of the electrodes is fully spin-polarized at the Fermi level (i.e.,~$P_L=1$ or~$P_R=1$). Interestingly, while in the parallel magnetic configuration $\mathcal{S}_\text{S}$ is always positive as a function of $P_L$ and $P_R$ (for  other parameters fixed), in the antiparallel configuration both positive and negative spin thermopower can be observed. Moreover, in the parameter space of $P_L$ and $P_R$,  a clear transition between such two regimes can be seen, which  in Fig.~\ref{Fig:Fig16}(h) is marked by a dashed line representing $\mathcal{S}_\text{S}=0$. The origin of this behavior can be explained by analyzing the competition between spin-up and spin-down channels in thermally stimulated spin transport, as already discussed above.

\section{\label{Sec:Conclusions}Summary and conclusions}

We analyzed the influence of magnetic anisotropy on spin-dependent thermoelectric effects in a nanoscopic system in the linear-response regime. In particular, a magnetic tunnel junction with a large-spin impurity incorporated into the tunneling barrier was considered  as an example. Using the approach based on a master equation,  we derived kinetic coefficients, Eqs.~(\ref{Eq:L00})-(\ref{Eq:L12_L21}), that relate charge, spin and heat currents to electrical, spin and thermal biases, Eq.~(\ref{Eq:L_matrix_def}). Knowledge of these  coefficients, in turn, allowed for finding quantities characterizing the spin-dependent thermoelectric response of the system, like charge, spin and thermal conductances, as well as both conventional and spin thermopowers.

We begun with considering the case of a bare junction, i.e., a  junction with no impurity, which was a reference point for further discussion. Then, the impurity was included and its growing impact  on transport characteristics of the junction was investigated by increasing the interaction between the impurity spin and tunneling electrons. This interaction leads to  scattering  of electrons traversing the junction, so that electrons can effectively exchange both angular momentum and energy with the impurity. Importantly, comparing three distinctive cases: an isotropic spin impurity with an anisotropic one of the `easy-axis' and `easy-plane' types, we showed that the uniaxial magnetic anisotropy is of key importance for such processes. Coupling to the impurity provides effectively additional transport channels, which in the presence of the uniaxial anisotropy are progressively activated with increasing temperature. This results in an increase in electrical, spin and thermal conductances. A more peculiar behavior  is observed in the case of thermopowers. The conventional thermopower is negative and its absolute value becomes increased in both parallel and antiparallel magnetic configurations. On the other hand, the spin thermopower is negative in the antiparallel configuration, with its absolute value diminished, while in the parallel magnetic configuration the spin thermopower  can change its sign, which generally stems from the competition between the spin-up and spin-down channels in thermally-induced spin transport. As the uniaxial spin anisotropy is usually accompanied by the transverse component, we investigated its impact on the transport and thermoelectric coefficients for an impurity with integer and half-integer spin numbers, finding that an especially profound effect can be seen in the former case.

Finally, we considered the transport characteristics of the system in the limit when the charge transfer through the junction is blocked. It was shown that owing to  the presence of the spin impurity, transport of spin and heat is still feasible in such a case. In principle, the impurity can be used as an intermediate reservoir of angular momentum and energy, to/from which these two quantities can be added/subtracted by electrons in single-electrode tunneling processes.
If such processes for one electrode result in delivering spin and energy to the impurity, while for the other electrode they lead to  transfer of the accumulated spin and energy from the impurity to this electrode, one gets a net flow of spin and energy across the junction. However, we note that energy can be transferred between the electrodes only if the impurity is capable of storing it, that is when transfer of energy is associated with transfer of angular momentum, which takes place only when the impurity exhibits uniaxial magnetic anisotropy. Interestingly, the system is then thermoelectrically responsive, showing a finite spin thermopower. The maximal attainable value of the spin thermopower at a given temperature depends largely on the spin-polarization  of the electrodes. Moreover, in the antiparallel configuration, the sign of the spin thermopower can be tuned by a proper choice of the spin polarization parameters.
We predict that at low temperatures, $T\lesssim\text{ZFS}$, the magnitude of spin thermopowers induced in this way  can in some conditions (e.g., for an integer spin in the parallel magnetic configuration) exceed by several orders those observed for a bare junction.


\acknowledgments

This work was supported by the National Science Center in Poland as the Project No. DEC-2012/04/A/ST3/00372.
\mbox{M.\,M.} acknowledges support from  the Alexander von Humboldt Foundation.



%


\end{document}